\documentstyle[epsf,preprint,aps]{revtex}
\tightenlines
\def\Journal#1#2#3#4{{#1} {\bf #2}, #3 (#4)}
\def\AP{{\em Ann. Phys.}}
\def\JPA{{\em J. Phys.} A}
\def\JPC{{\em J. Phys.} C}

\def\NPB{{\em Nucl. Phys.} B}

\def\PLB{{\em Phys. Lett.} B}
\def\PRL{\em Phys. Rev. Lett.}

\def\PRA{{\em Phys. Rev.} A}
\def\PRB{{\em Phys. Rev.} B}
\def\PRD{{\em Phys. Rev.} D}
\def\PRE{{\em Phys. Rev.} E}

\def\RMP{{\em Rev. Mod. Phys.}}
\def\PREPC{{\em Phys. Rep.} C}

\def\PHA{{\em Physica} A}
\def\PHY{{\em Physics}}

\newcommand{\be}{\begin{equation}}
\newcommand{\ee}{\end{equation}}
\newcommand{\bea}{\begin{eqnarray}}
\newcommand{\eea}{\end{eqnarray}}

\newcommand{\hf} {{1\over2}}
\newcommand{\nonu}{\nonumber\\}
\def\Dk{\Delta k}

\def\trp{{\rm tr'}}

\begin{document}
\title{Global Renormalization Group}

\author{Jean Alexandre\thanks{alexandr@lpt1.u-strasbg.fr}}
\address{Laboratory of Theoretical Physics, Louis Pasteur University\\
3 rue de l'Universit\'e 67087 Strasbourg, Cedex, France}

\author{Vincenzo Branchina\thanks{branchin@lpt1.u-strasbg.fr}}
\address{Laboratory of Theoretical Physics, Louis Pasteur University\\
3 rue de l'Universit\'e 67087 Strasbourg, Cedex, France}

\author{Janos Polonyi\thanks{polonyi@fresnel.u-strasbg.fr}}
\address{Laboratory of Theoretical Physics, Louis Pasteur University\\
3 rue de l'Universit\'e 67087 Strasbourg, Cedex, France\\
and\\
Department of Atomic Physics, L. E\"otv\"os University\\
Puskin u. 5-7 1088 Budapest, Hungary}

\date{\today}
\maketitle
\begin{abstract}
The motivation and the challenge in applying the renormalization group 
for systems with several scaling regimes is briefly outlined. The four 
dimensional $\phi^4$ model serves as an example where a nontrivial low 
energy scaling regime is identified in the vicinity of the spinodal 
instability region. It is pointed out that the effective theory defined
in the vicinity of the spinodal instability offers an amplification 
mechanism, a precursor of the condensation that can be
used to explore nonuniversal forces at high energies.
\end{abstract}

\section{Introduction}
The idea of the renormalization group \cite{kadanoff}, \cite{wilsrg}
is to view the different interactions
in a hierarchical manner, by building up the more complex systems from
their "elementary" constituents. The renormalization group should describe 
the manner the elementary particles rearrange themselves in forming 
the composite particles captured by the detectors with given resolution,
the complex structure of ordinary matter and finally 
the transition from the micro to the macroscopic physics.
Due to the obvious technical difficulties of such an ambitious
project the realization of this idea is severely restricted. What is
usually achieved by the help of different analytical approximation methods 
is to analyze the dependence on the observational scale in a scaling
regime where the evolution equations are linearizable or at least
perturbative. The result is the renormalized trajectory, the scale
dependence of the coupling constants in a regime that is dominated
by a given interaction. One may call this a local analysis of the 
renormalization group flow, performed in the individual scaling regimes
that usually but not necessarily agree with the vicinities of the fixed points.
A number of important results have been derived in this manner.

The real challenge, we believe, is to describe the transmutation of one 
set of scaling laws into another one as the scale of the observation is 
changed. This requires the construction of the renormalized trajectory 
connecting different scaling regimes. Such a manifestly nonperturbative
phenomenon will be studied in this paper in the case of a simple model
with two scaling regimes, the single component $\phi^4$ scalar field
theory in the spontaneously broken phase. This model supports an
asymptotic UV scaling regime well above the particle mass and an
intermediate energy scaling regime at the onset of the spinodal
instability. Our goal is to connect the two scaling regimes and the
global reconstruction of the renormalized trajectory. A preliminary
account of our results has already been given in ref. \cite{eger}.

The possibility of the the onset of a new scaling law has been raised
by the introduction of the "dangerously irrelevant parameters" \cite{fischer}.
The usual classification of the operators in a scaling regime 
with respect their response to the change of the scale is based
on the perturbation expansion and assumes the regular behavior of the
system in the weak coupling regime. If the evolution equations
develop singular behavior in the weak coupling limit then it may happen
that a perturbatively irrelevant coupling constant experiences new, 
nonperturbative scale dependence.
This mechanism can be identified in different systems with 
condensates \cite{senben}, \cite{enzo}, \cite{brmopo}, where
the singular dependence of the saddle points in the coupling constants
provides the mechanism to turn a perturbatively irrelevant coupling 
constant into an important one. It is conjectured in this paper that the
condensation might serve as a quite general mechanism to generate
new, important coupling constants and the dynamical renormalization
group that addresses the question of the scale dependence in 
non-equilibrium, time-dependent phenomena is well suited to the
study of such a question. 

We rely in this work on the renormalization group realized by 
a sharp cutoff in momentum space, in the framework of the gradient
expansion for the action. Such a realization of the
cutoff renders the systematic gradient expansion questionable,
so we restrict ourself to the lowest order, local-potential
approximation \cite{wh}, \cite{locpot}. But it should be mentioned that the usual remedies
of the problem, the use of a smooth cutoff or periodic Brillouin
zone are not compatible with the loop expansion for systems with a
condensate at finite-momentum scales. In fact, though the 
the loop expansion produces the action of the effective theory as 
a power series in $\hbar$, the saddle point, the minimum itself is not 
necessarily a polynom of $\hbar$. In order to preserve 
$\hbar$ as a small parameter, all of the stable modes have to be eliminated in 
the loop expansion before we compute the contributions of the unstable modes. 
Such a successive elimination of the degrees of freedom introduces 
automatically the strategy of the renormalization group with sharp cutoff,
and renders the smooth cutoff regularization inconsistent with the loop
expansion.

The organization of the paper is the following.
The presence of several scaling regimes is pointed out for the Theory of Everything
and for the BCS ground state in Section 2. The role the Bose condensation
plays in generating new scaling laws is elucidated in Section 3. The infinitesimal,
renormalization group equation is derived for the $\phi^4$ model
in Section 3. The possibility of a new scaling regime in the vicinity of the
spinodal instability is pointed out in Section 4. The numerical results for
the solution of the renormalization group equation are
presented in Section 5, in the case of the four dimensional $\phi^4$ model.
Finally, Section 6 contains the Summary.

\section{Hidden Coupling Constant}
The most fundamental and at the same time the most complex appearance of the multiple 
scaling laws can be found in the renormalized trajectory of the Theory of Everything, TOE. 
The simplest
procedure to describe the manner the TOE gives rise a chain of lower energy effective
theories is the so-called matching. This method where we match two theories at their
crossover scale can not give an account of the change of the scaling laws in a dynamical
manner, because different sets of coupling constants are used at the two sides
of the crossover. Instead, one should follow the original strategy of the Wilson-Kadanoff
blocking, and put all coupling constants in the lagrangian from the very beginning 
that will later be generated by the blocking. Thus the coupling constant space of 
the TOE should contain not only the renormalizable parameters but any 
coupling constant we ever need in Physics. For example the 
quark-gluon vertex or a coupling constant of the Hubbard model of the condensed 
matter physics have to be considered as complicated composite operators 
in terms of the fundamental particles of the TOE. 

\begin{figure}
	\epsfxsize=10cm
	\centerline{\epsffile{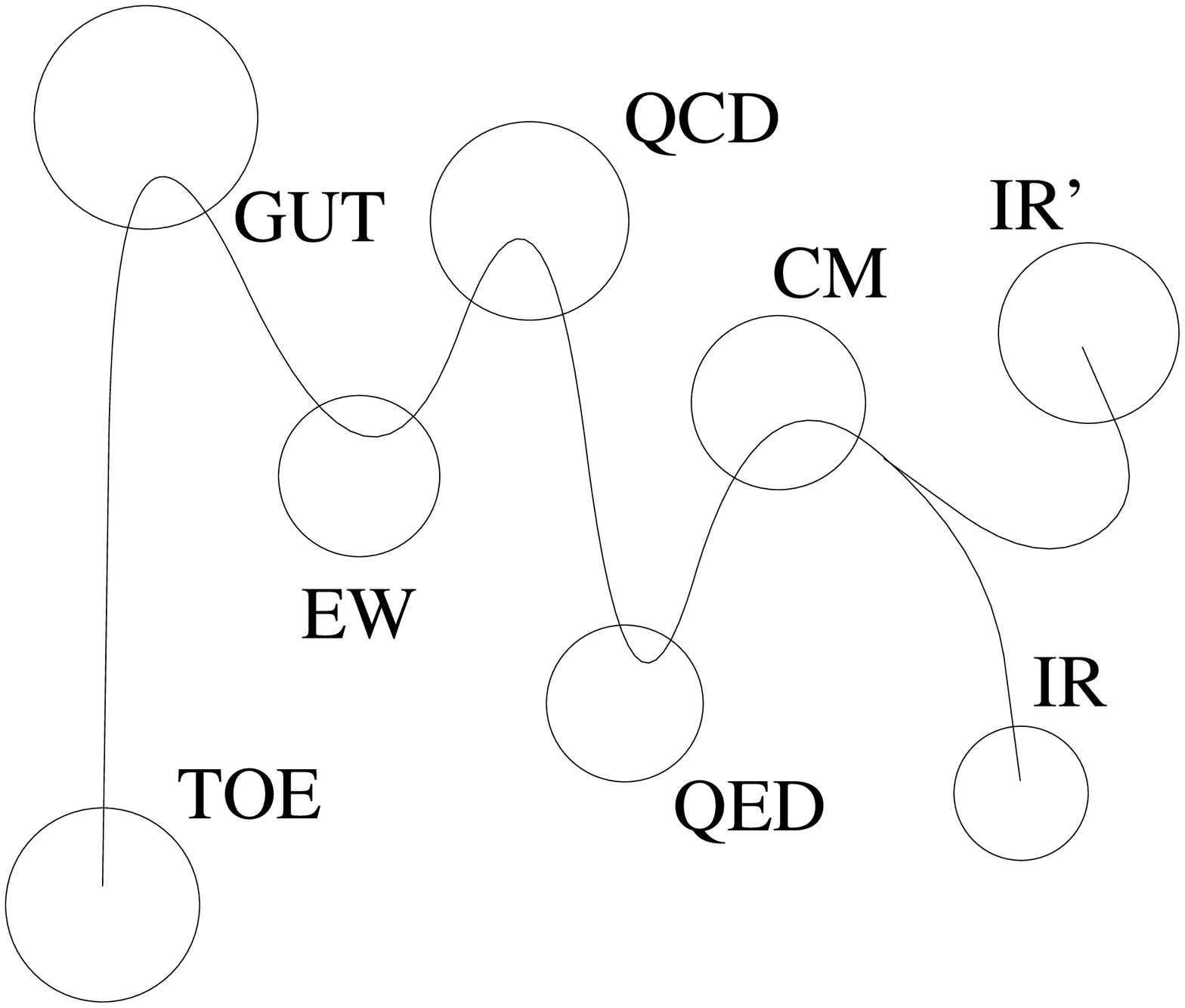}}
\caption{The renormalized trajectory of the Theory of Everything (TOE).
It passes by the fixed points of the Grand Unified Models (GUT), the
unified Electro-Weak theory (EW), the strong interactions (QCD), the
electromagnetic interactions (QED) certain fixed points of the Solid State
and Condensed Matter Physics (CM) and finally approaches the ultimate IR
fixed point. The trajectory may be influenced by the environment
and reach different thermodynamical phases in the IR regime.\label{rgtoe}}
\end{figure}

The renormalized trajectory, depicted schematically in Fig. \ref{rgtoe}, 
approaches several fixed points in its way towards the infrared limit.
To understand this better consider the increase of the observational energy in the 
regime 1-60GeV. The evolution of the coupling constants receive their dominant
contributions from the strong interactions, from the radiative corrections
of QCD, and the renormalized trajectory is in the scaling regime of QCD.
Had our world contained the strong interactions only, the renormalized
trajectory would have converged to the fixed point of QCD with increasing
energy. But the weak interactions become important as we reach the 
characteristic energy of the Electro-Weak theory, and the renormalized trajectory
turns away from the QCD fixed point. This happens because the running coupling 
constants of the non-renormalizable quark vertices 
generated by the exchange of the intermediate vector bosons increase with
the energy. They saturate at the crossover between the strong and the
electro-weak interactions where the guidance of the evolution is taken over 
by the exchange of the W and the Z bosons. 
In a similar manner, the fixed points of all the other, renormalizable,
effective theories are approached by the renormalized trajectory, but the 
higher-energy
processes always prevent convergence as the energy is increased, except
at the last fixed point, at the TOE. In the regime of Solid State Physics,
we can influence of the evolution of the running coupling constants by
the environmental variables, such as the temperature or chemical potentials \cite{carg}.
In this manner the renormalized trajectory may bifurcate and follow
different path in different environments and finally arrive in different
thermodynamical phases at the infrared fixed point.

Facing such a complex system, the usual argument about universality appears
as an oversimplification. In fact, at each scaling regime we classify the
operator algebra of the model in a local manner according to 
the appropriate scaling laws. It may happen that an operator possesses different local 
classifications, and it is found relevant at one scaling regimes but becomes irrelevant at 
another one. The importance or unimportance of such a coupling constant must be
decided in a scaling-regime-independent, global manner. 

A simpler example where the global behavior is important can be found in QED 
containing the electron and a heavy pointlike particle with charge +Z playing the 
role of a nucleus. Suppose that certain environment variables, such as the temperature 
and the baryon chemical potential are chosen in such a manner that the vacuum 
is a solid state lattice in the superconducting phase. We can distinguish two
asymptotic scaling regimes in this model:

\begin{itemize}
\item {\it Asymptotic UV scaling:} At energies above the nucleus mass the evolution
equation is given in terms of the minimal coupling vertices and the vacuum can be
considered perturbative. As indicated above we do not require the 
existence of a fixed point for the identification of the asymptotic
scaling laws. In this manner we can ignore the possible problems arising from the
non-asymptotically free character of QED, and consider the scaling laws only up to 
the UV Landau pole. The relevant and the marginal operators are the usual 
renormalizable ones, and are given in the framework of the perturbation expansion by a
power-counting argument. The size of the scaling regime is limited by the UV cutoff or the
Landau pole.

\item {\it Asymptotic IR scaling:} For energies below the scale of eV the collective
phenomena of the solid state lattice dominate the scaling laws. The inhomogeneity of
the vacuum is a key element. No systematic classification of the scaling 
operators is known, but due to the massless acoustic phonons, the existence
of nontrivial relevant or marginal operators can not be excluded. The lower edge
of the scaling regime is limited by the loss of quantum coherence.
\end{itemize}

\begin{figure}
	\epsfxsize=10cm
	\centerline{\epsffile{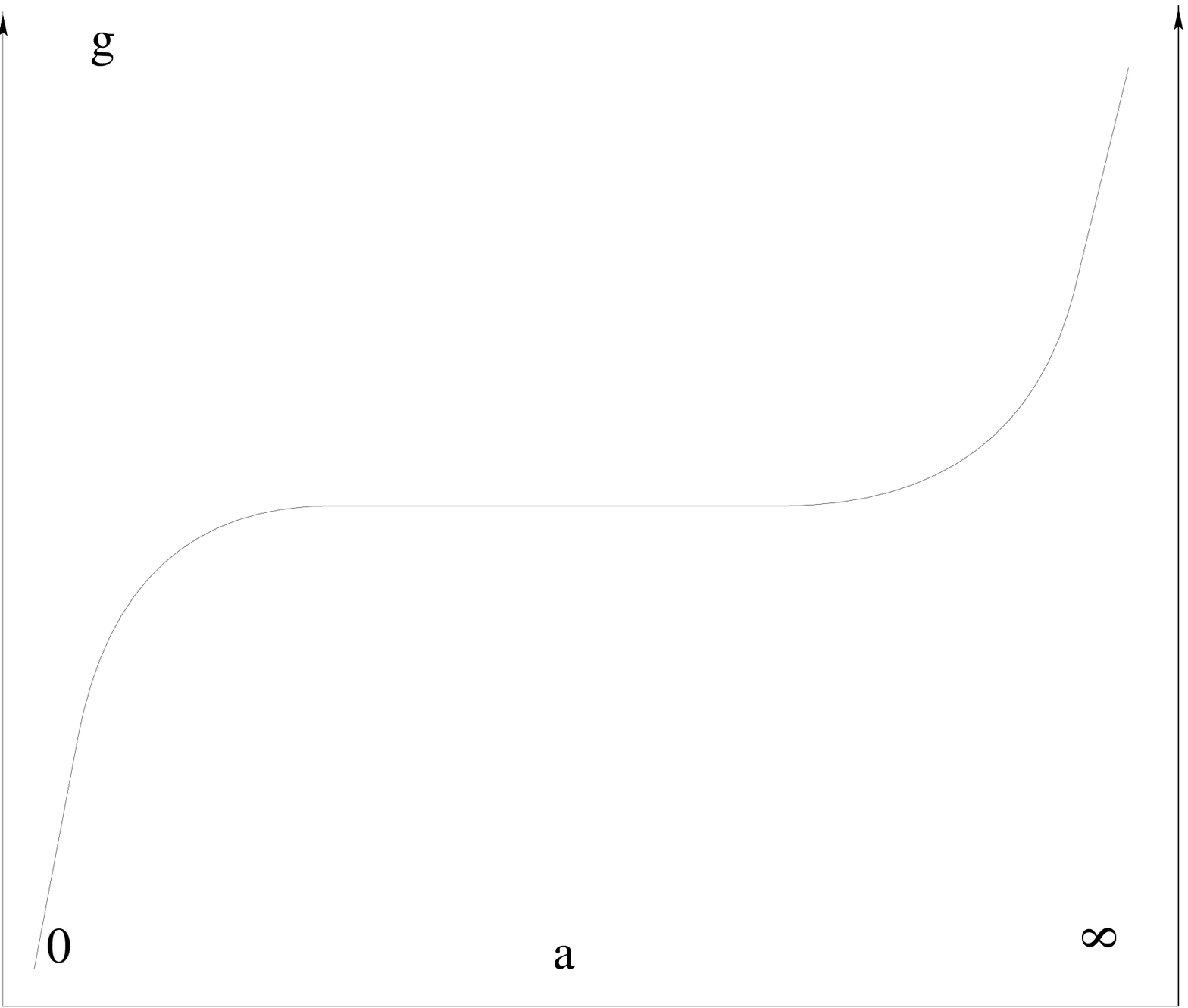}}
\centerline{(a)}
	\epsfxsize=10cm
	\centerline{\epsffile{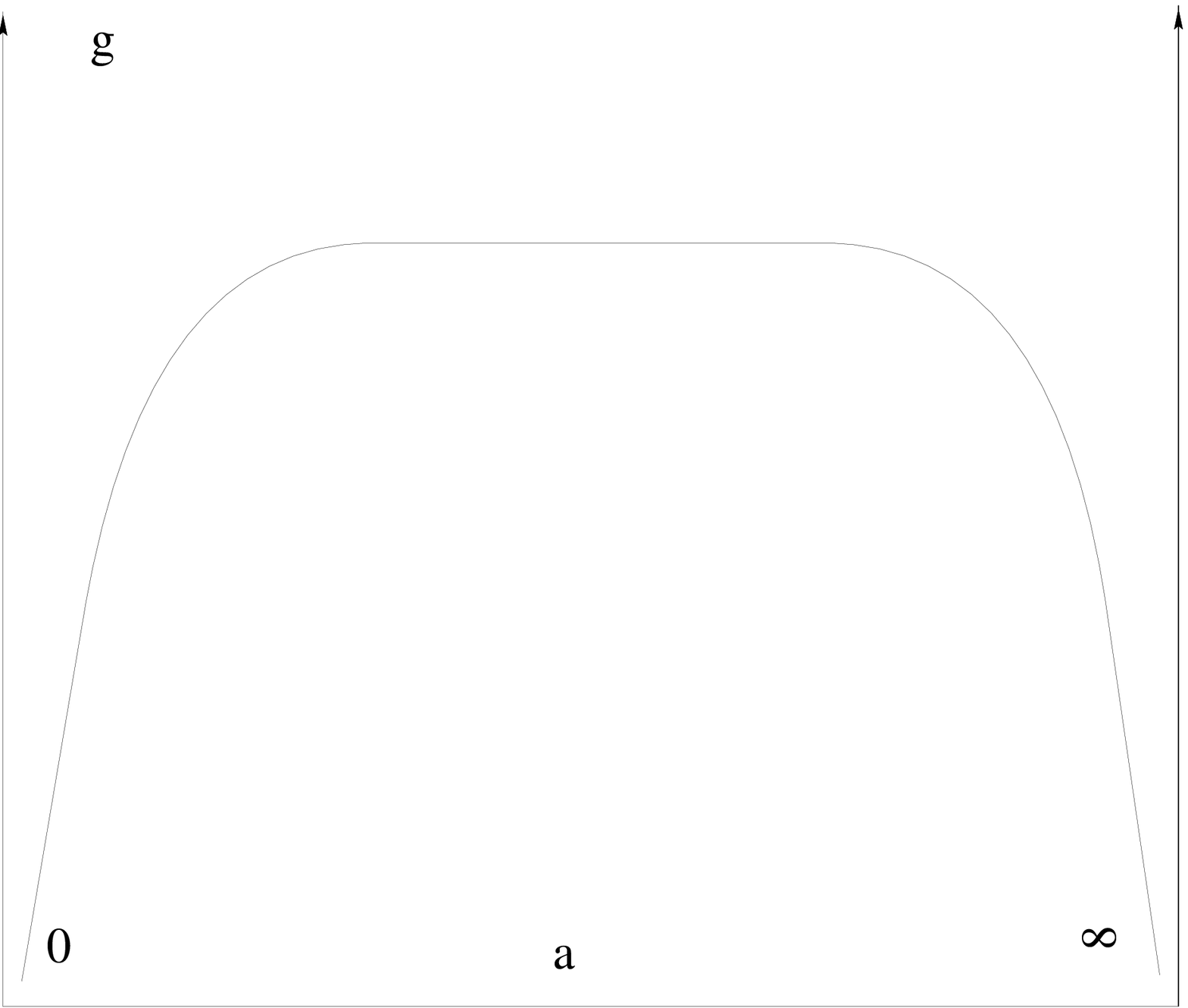}}
\centerline{(b)}
	\epsfxsize=10cm
	\centerline{\epsffile{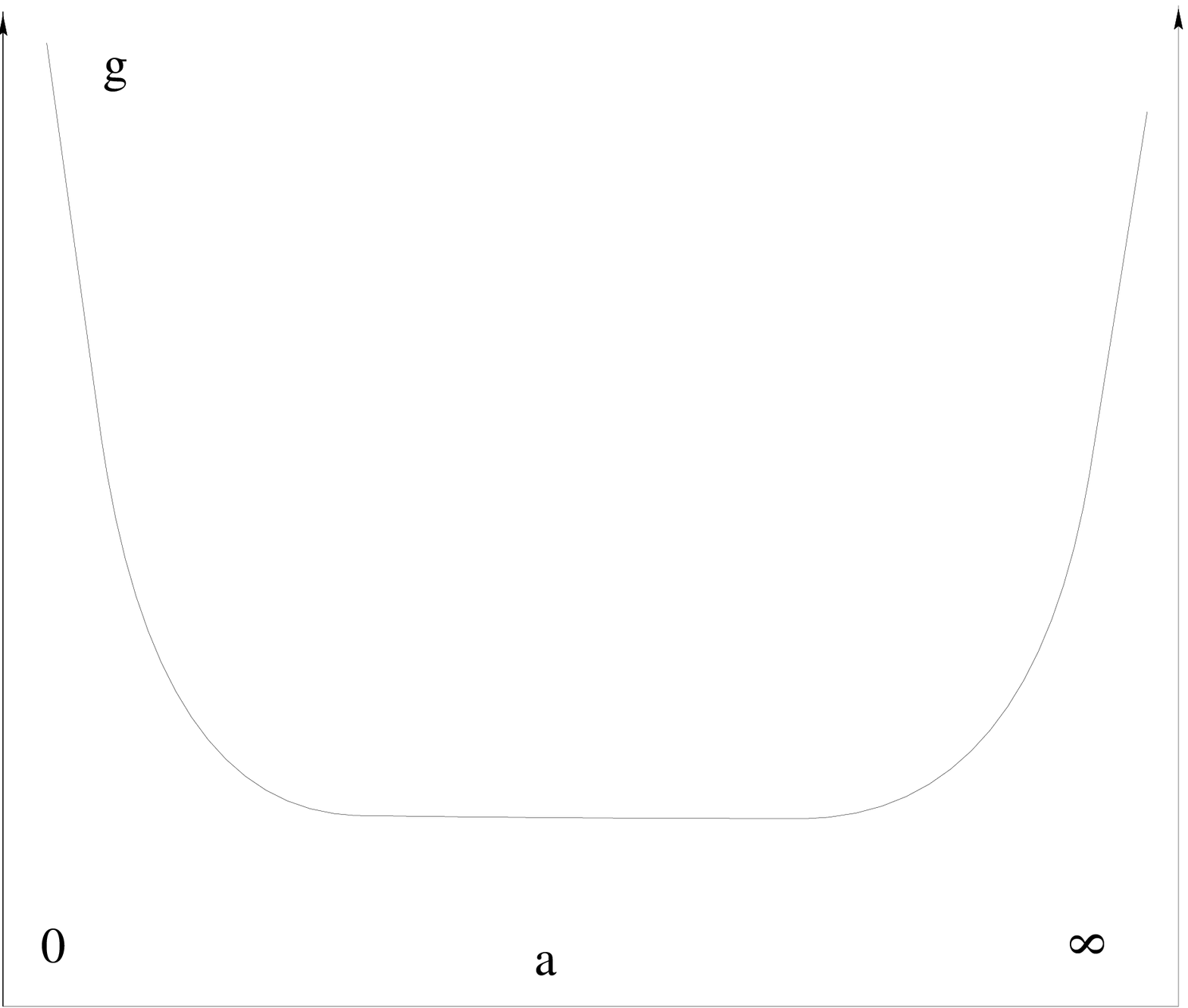}}
\centerline{(c)}
	\epsfxsize=10cm
	\centerline{\epsffile{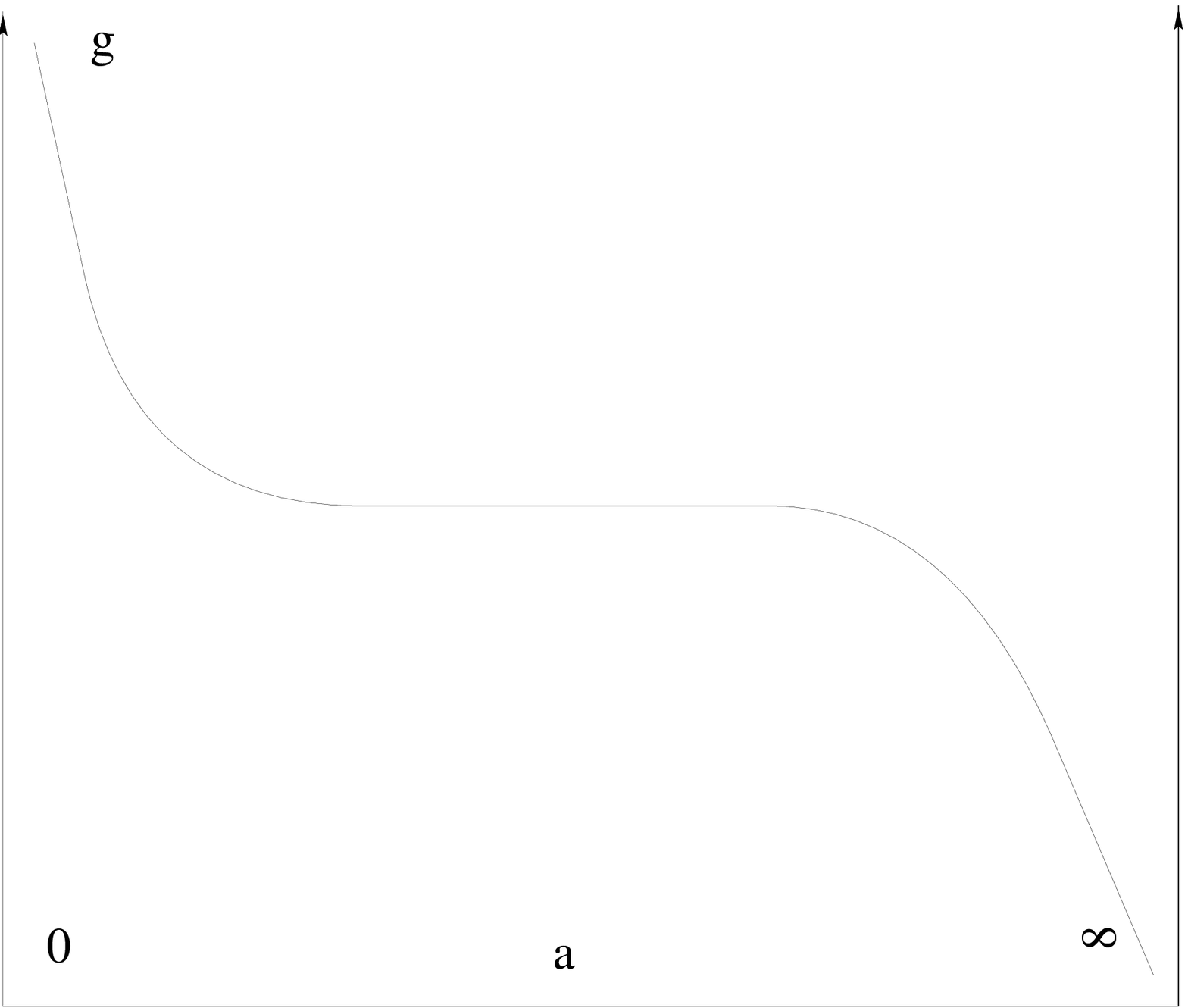}}
\centerline{(d)}
\caption{The qualitative dependence of the running coupling constant
of Table 1 as the function of the cutoff, $a=2\pi/\Lambda$. The asymptotic 
UV and IR scaling regimes are shown. The coupling constant is supposed to be
constant in between for simplicity.\label{qedccp}}
\end{figure}

The asymptotic scaling regimes can be extended by embedding the model into a 
higher-energy, more fundamental, renormalizable theory and by approaching 
absolute-zero temperature. These asymptotic scaling laws are important since their
relevant coupling constants may influence the physics in a strong manner.
The four different possibilities with respect to these scaling laws are shown in Table \ref{qedcc}
and in Fig. \ref{qedccp} for coupling constants whose dimension is removed
by the cutoff, $\Lambda$. The electron mass is a relevant parameter of the QED lagrangian
and remains relevant for the solid state lattice, too. The muon mass is relevant in the 
UV scaling regime but the muon-induced processes
are overwhelmed by the electron-induced ones at low energy and $m_\mu$ becomes
irrelevant in the IR. The four-fermion interaction is, at the same time,
non-renormalizable and represents the driving force to the BCS superconducting 
phase; it is marginal in the IR regime \cite{shankar}, \cite{polch}. Recently other indications of 
the deviation from the Fermi-liquid behavior resulting from relevant or marginal operators
of the IR regime have been found for high-$T_c$ cuprates \cite{htc}, as well.
Finally the six-fermion or any higher order vertex is irrelevant in the IR regime because 
its effect is reduces to the multiple application
of the four-fermion vertex. The four-fermion interaction plays a special role:
On the one hand, it is usually left out from the microscopic lagrangian because 
it is suppressed in the UV scaling regime. On the other hand, it has a key role
at the IR regime in controlling the attraction between the electrons. 

\begin{figure}
	\epsfxsize=10cm
	\centerline{\epsffile{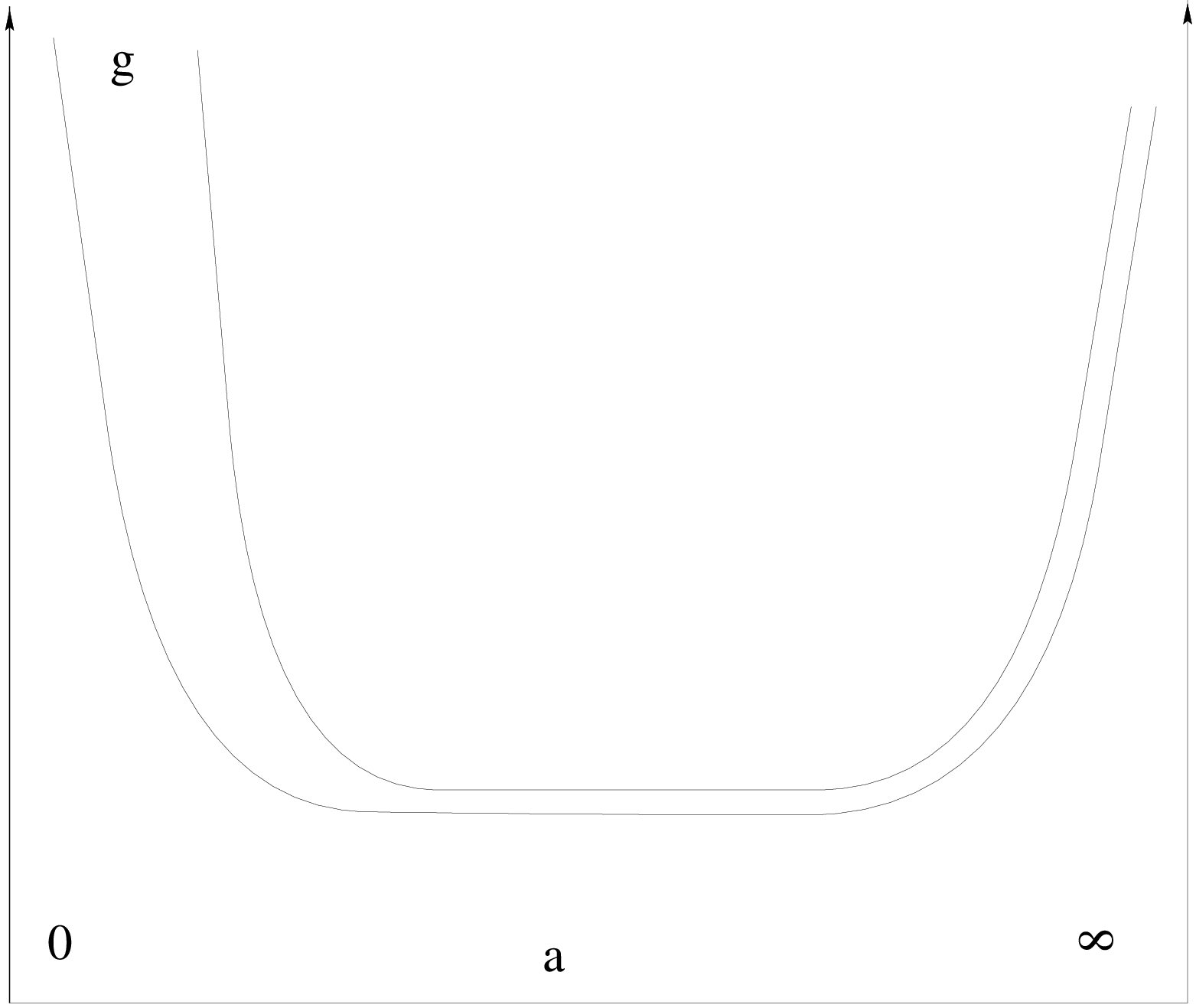}}
\centerline{(a)}
	\epsfxsize=10cm
	\centerline{\epsffile{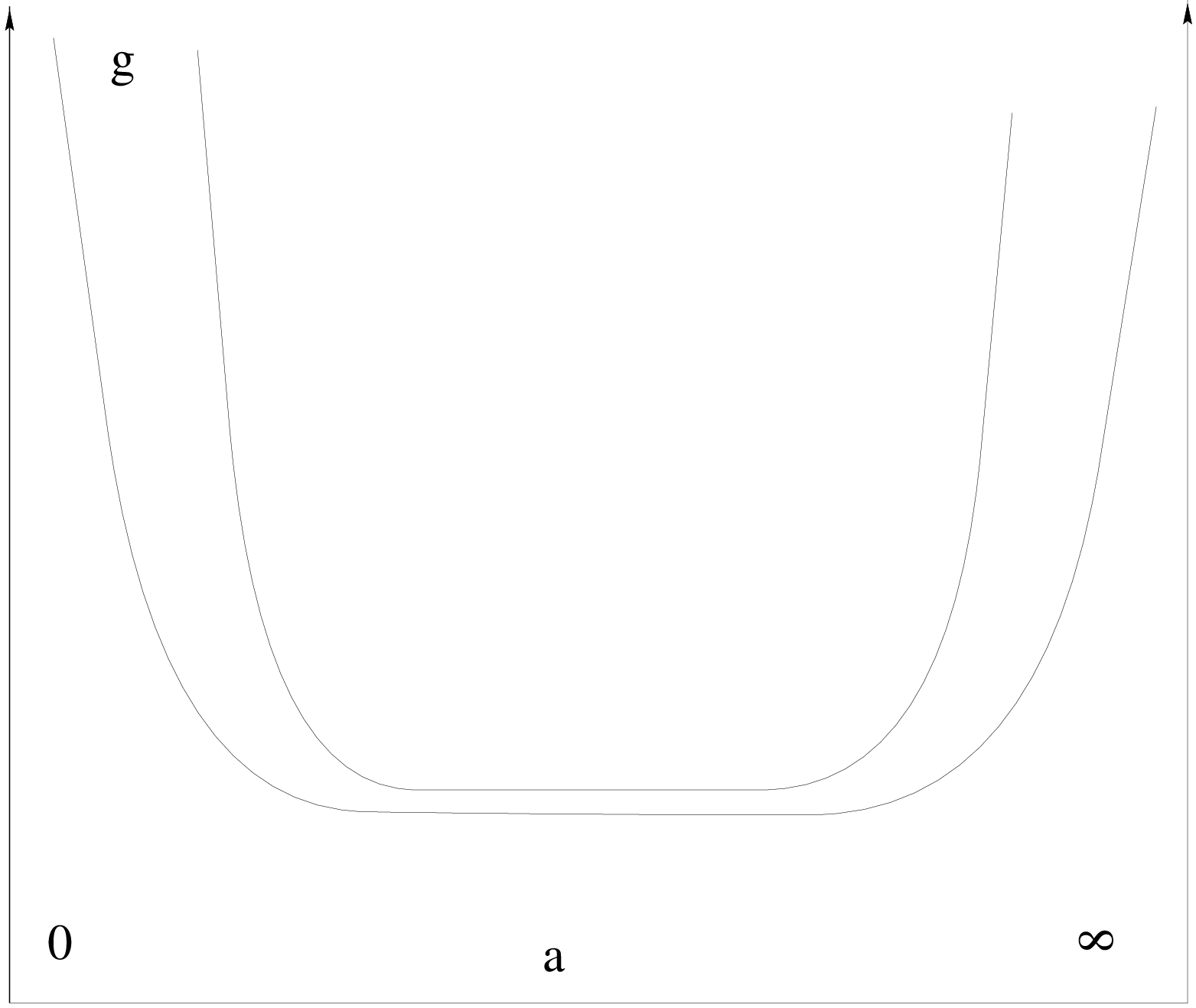}}
\centerline{(b)}
\caption{Two possible behaviors of a nonrenormalizable coupling constant what is
relevant at the low energy scaling regime, case (c) in Table 1. Each plot shows
two curves what belong to initial conditions at the cutoff what differ in the
value of the irrelevant coupling constant only.\label{qedh}}
\end{figure}

The qualitative behavior shown in Fig. \ref{qedccp}c raises the 
following possibility. The suppression of the irrelevant coupling constants in the UV
scaling regime is used to explain the universal behavior of the models.
It is certainly correct to expect that the renormalized trajectories
whose initial conditions in the ultraviolet differ only in the values of the
irrelevant coupling constants approach each other as we move towards
the infrared direction. But if there is another scaling regime where an operator
that was irrelevant in the UV scaling regime turns out to be relevant, then
the resulting amplification process may undo the suppression at the UV. May one
find a "hidden coupling constant" in this manner that has to be put into the microscopic
action of the UV regime but influences the dynamics only in the IR ? The answer to
this question is nontrivial even if we can identify a nonrenormalizable 
operator that is relevant in the low-energy scaling regime.
In fact, it may happen that the increasing value of this coupling constant
during the lowering of the observational energy happens to be independent of its initial
value at the UV cutoff. This possibility is shown schematically in Fig. \ref{qedh}a.
The UV scaling laws suppress the dependence on the intial values of the
nonrenormalizable coupling constants. If the evolution equations have no
instability or other nonanalytic features at low energies, then this suppressed
sensitivity and universality is observed down to zero energy. 
In Fig. \ref{qedh}b the value of the coupling constant at the cutoff becomes 
numerically negligible at intermediate energies according to universality. 
However the low-energy scaling laws amplify the numerically-small nonuniversal 
value. Dependending on the critical exponents and the size of the scaling regimes
the amplification of the sensitivity at the IR side may be comparable or even stronger 
than the suppression in the UV regime. This is a hidden coupling constant because it is
undetectable at finite energies nevertheless its small value influences the IR
physics in a nonuniversal manner. By rearranging a longer IR evolution 
we can in principle uncover the presence of nonuniversal interactions at higher
energies.

It was pointed out that a relevant operator of the IR 
scaling regime may change the ground state either by generating bound 
states that condense or by driving the system to strong couplings until the 
growth of the coupling constants is cut off by other quantum effects \cite{polch}.
Within the framework of the saddle-point expansion these two possibilities
coincide, and the issue of the hidden coupling-constant is that the growing coupling 
constants might be slowed down in a manner that contains information about the 
nonuniversal interactions at the microscopic scale. In order to clarify this question, 
namely the sensitivity of the IR end of the renormalized trajectory on the UV 
initial conditions, we must obtain and solve the renormalization group equations 
for sufficiently many coupling constants, globally.

The coexistence of several scaling regimes have already been studied in 
Condensed Matter Physics,
where competing interactions are represented by the possibility of approaching
different IR fixed points \cite{cubic}. The difference between such systems and the TOE is that
only one of the possible scaling laws are realized in the former case. On the contrary,
the scaling regimes occur at different energy scales and the system visits each of them
sequentially in the TOE.

The renormalized trajectory of TOE sketched in Fig. \ref{rgtoe} reflects the
usual conflict between the "fundamental" and "applied" physics. The fundamental,
microscopic parameters should be determined by the help of the renormalization
conditions imposed at the energy scale $\mu_{fund}$ that is within the scaling regime of the fundamental
interactions, and is far from the complexity entering at lower energies. So long as
the TOE is renormalizable or finite the perturbation expansion can be used
to show that the resulting renormalized trajectory is independent of the choice of
$\mu_{fund}$, and can in a natural manner be characterized by the coupling constants
observed at high energies. In this sense the fundamental laws of physics are determined 
by the high-energy experiments and the description of the lower-energy, complex systems 
requires "only" the capacity to apply the fundamental laws in a complicated situation.

But the fallacy of this view is clear: In the absence of hidden coupling constants
where the low-energy effective-parameters are determined in an autonomous manner,
there is no need for the precise measurement of the high energy parameters
in order to reproduce the low-energy physics. In fact, according to the renormalized
perturbation expansion and universality, the bare coupling constants are characterized
in a unique manner by the renormalized ones defined at low energies, 
$\mu_{compl}<<\mu_{fund}$. When hidden coupling constants
are present then the precision of the measurement of the high-energy parameters 
required to predict a low-energy phenomenon with a reasonable accuracy might render the
global determination of the renormalized trajectory illusory. In other words,
the specification of the initial conditions with a reasonable accuracy
might be unimportant for the nonlinear evolution, similarly to
the nonintegrable chaotic systems. The challenge lies in comprehending the matching of 
the different scaling islands of Fig. \ref{rgtoe}.

\section{Renormalization and Condensation}
In the perturbatively-implemented renormalization group we have to 
assume that the renormalized trajectory does not leave the vicinity of the
Gaussian fixed point. If there is a small parameter assuring this, like 
in $4-\epsilon$ dimensions, then the set of the relevant 
or marginal operators is the same at each scaling regime,
and the global behavior of the renormalized trajectory contains nothing new
compared to the local analysis performed at the individual scaling regimes. If the
renormalized trajectory explores the nonperturbative regions then the global
analysis is very interesting, but we loose the general, perturbative characterization of
the flow. A window of 
opportunity for the analytical studies opens if the nonperturbative features of the
evolution can be reproduced by the only known, systematic, analytical, nonperturbative 
method, the semiclassical expansion. 

A nontrivial saddle point corresponds to a coherent state formed by the condensate
of bosons. The importance of this condensate depends on the multiplicity of the
coherent state(s). We shall demonstrate this from three different points of view.

\begin{itemize}
\item {\it Euclidean Quantum Field Theory for the vacuum.}
In the semiclassical solution to the vacuum with 
ferromagnetic condensate only the lowest lying excitation level with vanishing
momentum is populated macroscopically.
\footnote{The degeneracy of the vacuum with spontaneously broken symmetries
does not change the picture since there is only one vacuum in each
"world" containing the states connected by local operators.}
When the system is placed in a finite geometry then the lowest lying state might 
become inhomogeneous; nevertheless the particles still condense in a single mode. 
The impact of the condensate is stronger if the particles form coherent
states in a large number of modes. This is the case for solitons or instantons,
the localized saddle points with high entropy. Another, less studied example is given by
theories with higher-order derivative-terms in the action \cite{brmopo}. When their
coupling constant is properly chosen then particles with nonvanishing momentum 
may condense and generate an inhomogeneous vacuum. Since the density of modes with
momentum $p$ is proportional to $p^{d-1}$ in $d$ dimensions there are more particle modes
participating in the condensation at finite $p$ than at $p=0$.

\item {\it Real time dynamics.} The condensation is triggered by the wrong sign of the 
forces trying to restore the equilibrium position. There are always several unstable
modes which compete in the condensation when the system starts with the naive vacuum 
\cite{boytd}. Depending on the initial conditions different slower collective modes 
might be formed that distribute the energy and create time dependent condensates in 
several other particle modes. After sufficiently long times, the energy of the excited modes
get diffused by the friction terms over infinitely many modes, without creating
macroscopic population, and only the particles in the lowest lying mode remain condensed.
The large-amplitude, transient excitations above the true vacuum are described 
by a large number of modes with condensate. This is the realm of 
nucleation and the spinodal decomposition \cite{spinod}. 

The former refers to the 
metastablility of the vacuum which decays by large amplitude fluctuations.
The simplest is to assume that this takes place by the spontaneous formation 
of spherical droplets of the stable vacuum whose free energy is 
$F=4\pi R\sigma-4\pi R^3\Delta F/3$, where 
$\sigma$, $\Delta F$ and $R$ stand for the surface tension, the
free energy density difference betwen the metastable and the stable vacuum and the
droplet radius, respectively. 
These droplets extend over the whole volume if they are sufficiently large, 
$R>R_{cr}=\sigma/\Delta F$. The spinodal decomposition 
is observed when there is no more finite threshold for the instability and the 
infinitesimal fluctuations are enough to trigger the decay of the homogeneous 
vacuum. This can be recognized in the framework of the static 
description by the apparence of the negative eigenvalues for the
Euclidean propagator. The fast increase of the corresponding elementary
excitation amplitudes drives an inhomogeneous separation of two stable
values of the local dynamical variable. 

Returning to the long-time, low-energy excitations, they experience a single 
condensed mode; and the static nonperturbative condensate is simpler to describe.

\item {\it Saddle points of the renormalization group.}
When the renormalization group is implemented then we eliminate the modes in descending 
order in the energy, and we may encounter nontrivial saddle points in the way.
If we are interested in the effective theory for the low-energy fluctuations around
the true vacuum in systems without a localized saddle point, then the condensate
occurs only at the last mode, at the infrared fixed point. When the effective theory is 
sought for large amplitude excitations, then the eliminations of the modes in the 
presence of such a background field may induce saddle points for the blocking procedure
earlier, at finite scales. The nonperturbative contribution of these saddle points 
may modify the direction of the renormalization group flow in a substantial manner
\cite{hahad}. As an example of this mechanism, the emergence of an irrelevant coupling 
constant in the dynamics of domain walls has already been noted in the two dimensional 
nonlinear $O(2)$ model \cite{jasnow}.
\end{itemize}

In short, the saddle points of the blocking procedure occur with higher multiplicity
at finite energies, and may modify the scaling laws in a more substantial manner
than the saddle points of the static system in the true vacuum. Such effects
could be seen by the proper implementation of the dynamical renormalization 
group for large amplitude fluctuations. It is worthwhile noting that 
the spinodal instability region starts just at the vacuum in the presence 
of the Goldstone modes, and the saddle point contributions might influence the 
small quantum fluctuations around the true vacuum, as well \cite{albapo}.

\section{Renormalization Group Equation}
The study of the different scaling regimes requires the handling of a large number of
coupling constants which can be achieved by converting the evolution equations
referring to the individual coupling constants into a differential equation
for the generating function for the coupling constants, i.e. the blocked action. 
We shall work with the four-dimensional $\phi^4$ theory in Euclidean space-time.
The cutoff $k$ in the momentum space will be changed infinitesimally, $k\to k-\Dk$,
which generates a new small parameter, $\Dk/M$, where $M$ is a quantity of 
dimension of mass made up by the dimensional parameters of the model and the cutoff, $k$.
This small parameter will be used to show that the 
renormalization group equation derived in the one-loop approximation becomes exact
as $\Dk/M\to0$, because the higher loop contributions are suppressed by $\Dk/M$.

There are two limitations to bear in mind in turning this scheme into a feasible
algorithm. The first is that even thought there is a new small parameter to suppress
the higher loop contributions to the evolution equations the argument presupposes the
applicability of the loop expansion. Therefore the use of the resulting "exact" equation 
is questionable in the strong coupling region, beyond the validity of the loop expansion. 
The other limitation comes from the parametrization of the effective-action
functional, $S_k[\phi]$, with the cutoff $k$. It is usually done my relying on 
the gradient expansion,
\be
S_k[\phi]=\int d^dx\left[Z_k(\phi(x))\hf(\partial_\mu\phi(x))^2+U_k(\phi(x))\right]
+O(\partial^4).
\label{lagr}
\ee
The assumption that terms with higher order derivatives are less important is 
equivalent with the belief that the action is a local functional. This is 
reasonable at high energies because all of the relevant operators of the short distance 
scaling laws should be local. But there is no reason to exclude operators
which are nonlocal at the scale of the cutoff at low-energy, scaling regimes, and
these terms of the action may create nonlocal effects. By holding to the
assumption of locality we shall set 
\be
Z_k(\phi)=1\label{zegy}
\ee
in this work. 

We eliminate the modes with momentum $k-\Dk<p<k$ and find the blocked action
\be
e^{-{1\over\hbar}S_{k-\Dk}[\phi]}=\int D[\phi']e^{-{1\over\hbar}S_k[\phi+\phi']}
\label{brg}
\ee
The Fourier transforms of the variable $\phi(p)$ and $\phi'(p)$ are nonvanishing for 
$p<k-\Dk$ and $k-\Dk<p<k$, respectively. The functional integration is carried
out in the framework of the loop expansion,
\be
S_{k-\Dk}[\phi]=S_k[\phi+\phi_0']+{\hbar\over2}\trp\ln\delta^2S+O(\hbar^2),\label{elim}
\ee
where $\trp$ denotes the summation within the shell $[k-\Dk,k]$.
The positive semi-definite operator 
\be
\delta^2S(x,y)={\delta^2S_k[\phi+\phi']
\over\delta\phi'(x)\delta\phi'(y)}_{\big\vert\phi'(x)=\phi_0'(x)}
\label{inpr}
\ee
and the saddle point, $\phi_0'$,
\be
{\delta S_k[\phi+\phi']\over\delta\phi'(x)}_{\big\vert\phi'(x)=\phi_0'(x)}=0
\ee
are computed by keeping the background field, $\phi(x)$, fixed. Eq. (\ref{elim})
reduces to the Wegner-Houghton equation \cite{wh} when the saddle point is trivial,
$\phi_0'=0$.

Observe that the loop integration is made in a region of volume 
$\Omega_dk^{d-1}\Dk$ where
\be
\Omega_d={2^d\pi^{d/2}\over\Gamma(d/2)}.
\ee
The contributions $O(\hbar^n)$ are given in terms of $n$-fold loop integrals. 
So long as the integrands are bounded in the domain of integration the
contributions $O(\hbar^n)$ contain the multiplicative factor $(\Dk/M)^n$.
The integrands are the products the propagator, the inverse of (\ref{inpr})
evaluated within the subspace $k-\Dk<p<k$. If the restoring force acting on 
the fluctuations is nonvanishing then the propagator is bounded
and the higher loop contributions drop from (\ref{elim}) when
$\Dk/M\to0$. The formal argument showing the suppression of the higher
loop contribution to the renormalization group equation is presented in
the Appendix.

The only quantity needed in the local potential approximation, (\ref{lagr})-(\ref{zegy}), 
is the potential $U_k(\Phi)$ so it is natural to choose for its determination 
a homogeneous background field, $\phi(x)=\Phi$, for which the kinetic energy is vanishing. 
An additional bonus of this choice is that the saddle point is nontrivial, 
$\phi_0'\not=0$, just in the spinodal unstable phase.
Outside of the spinodal unstable phase one can set $\phi'_0=0$ what yields
\bea
e^{-S_{k-\Dk}[\Phi]}&=&e^{-L^dU_{k-\Dk}(\Phi)}\nonu
&=&e^{-S_k[\Phi+\phi_0']-{\hbar\over2}\trp\ln\delta^2S+O(\hbar^2)}\nonu
&=&e^{-L^dU_k(\Phi)-{\hbar\over2}L^d\int'{d^dp\over(2\pi)^d}
\ln\left[p^2+U_k''(\Phi)\right]+O(\hbar^2)},
\eea
where the plane wave elementary excitations of the homogeneous
vacuum were used to compute the trace in the third line
and $\int'd^4p=\int_{k-\Dk<p<k}d^4p$.
In this manner one arrives at the finite difference equation
\be
U_k(\Phi)-U_{k-\Dk}(\Phi)
=-{\hbar\over2}\int{d^dp\over(2\pi)^d}\ln\left[p^2+U_k''(\Phi)\right]
\left[1+O\left({\hbar\Dk\over M}\right)\right],\label{fdwh}
\ee
where
\be
U_k''(\Phi)={\partial^2U_k(\Phi)\over\partial\Phi^2}.
\ee
By taking the limit $\Dk\to0$ we find the differential equation
\be
k{\partial\over\partial k}U_k(\Phi)=-\hbar{\Omega_dk^d\over2(2\pi)^d}
\ln\left[k^2+U_k''(\Phi)\right].\label{rgde}
\ee
by the help of the polar coordinates in the momentum space.

The evolution equation, (\ref{rgde}), is actually the one-loop resummation of the perturbation
expansion. This can be seen by expanding in the non-Gaussian pieces of the potential,
\be
k{\partial\over\partial k}U_k(\Phi)=-\hbar{\Omega_dk^d\over2(2\pi)^d}\left[
\ln(k^2+m^2_k)+\sum_{n=1}^\infty{(-1)^{n+1}\over n}
\left({U_k''(\Phi)-m^2_k\over k^2+m^2_k}\right)^n\right],\label{lerg}
\ee
where $m^2(k)=U_k''(0)$. The evolution equation for the effective vertices at
zero momentum, i.e. the leading order of the gradient expansion is now the explicit 
sum of the one-loop graphs defined by the free propagator $G^{-1}(k)=k^2+m^2_k$. 
If we ignore the $k$-dependence in the right hand side then $U_0(\Phi)$ is the 
usual one-loop effective potential obtained in the bare or the renormalized 
perturbation expansion for 
\be
m^2(k)\to m^2_B~~~U_k''(\Phi)-m^2_k\to{\lambda_B\over2}\Phi^2,
\ee
or
\be
m^2(k)\to m^2_R~~~U_k''(\Phi)-m^2_k\to{\lambda_R\over2}\Phi^2,\label{reaf}
\ee
respectively.
The evolution generated by the renormalization group provides a partial
resummation of the perturbation expansion by piling up the effects of the
modes which have already been eliminated in the $k$-dependence of the
running coupling constants, $U_k(\Phi)$, in our case. In the usual renormalization
group method, which is based on the renormalized perturbation expansion, only the renormalizable
coupling constants, $\partial^2U_k(0)/\partial\Phi^2$, $\partial^4U_k(0)/\partial\Phi^4$, 
pile up the effects of the elimination. In solving the differential equation 
(\ref{lerg}) the nonrenormalizable coupling constants, i.e. the higher order derivatives
are evolved, as well. Such an extension of the usual scheme is necessary to find
the eventual hidden coupling constants.

It is easy to perform the direct numerical integration of (\ref{rgde}) with 
an initial condition $U_\Lambda(\Phi)$ given at $k=\Lambda$ towards the infrared direction
after sufficient attention is paid to the instabilities arising from the finite
$\Phi$ and $k$ resolution. However our goal is to arrive at a description of
the modification of the scaling laws which requires that the 
restoring force for the fluctuations, the argument of the logarithm function
in (\ref{rgde}), be vanishing at $k=k_0(\Phi)\not=0$. 
The singular behavior of the
differential equation made the numerical quadratures unstable and forced us to
resort to methods other than the discretization of the $\Phi$-space. Instead of
discretizing the variable $\Phi$ we truncated the Taylor expansion 
of $U_k(\phi)$ at finite order. 

To this end we introduce the coupling constants defined at $\Phi=\Phi_0$ by
\be
g_n(k)={\partial^nU_k(\Phi_0)\over\partial\Phi^n}.
\ee
The result is
\be
U_k(\Phi+\Phi_0)=\sum_n{g_n(k)\over n!}\Phi^n
\ee
The $\beta$-functions are defined as
\be
\beta_n=k{\partial\over\partial k}g_n(k)
={\partial^n\over\partial\Phi^n}k{\partial\over\partial k}
U_k(\Phi)_{\big\vert\Phi=\Phi_0},
\ee
where the analyticity of the potential in $k$ and $\Phi$ was assumed. They are
obtained by taking the successive derivatives of (\ref{rgde}),
\be
\beta_n=-\hbar{\Omega_dk^d\over2(2\pi)^d}{\cal P}_n(G_1,\cdots,G_{n+2}),
\ee
where
\be
G_n={g_n\over k^2+g_2}.
\ee
and
\be
{\cal P}_n={\partial^n\over\partial\Phi^n}\ln\left[k^2+U_k''(\Phi)\right]
\label{betader}
\ee
is a polynom of order $n/2$ in the variables $G_j$, $j=2,\cdots,n+2$. 

In order to exploit the simplification offered by the symmetry $U_k(-\Phi)=U_k(\Phi)$
we set $\Phi_0=0$, which cancels the odd vertices and yields for the even ones
\bea\label{pebfv}
{\cal P}_2&=&G_4,\nonu
{\cal P}_4&=&G_6-3G_4^2,\\
{\cal P}_6&=&G_8-15G_6G_4+30G_4^3,\cdots\nonumber
\eea
One can verify diagrammatically that the system of equations 
(\ref{betader})-(\ref{pebfv}) 
is a compact rewriting of the one-loop contribution to the beta functions of 
the effective vertices.

It is useful to obtain the evolution equations for the coupling constants 
whose dimension is removed by the cutoff, 
\bea
\tilde\beta_n&=&\tilde k{\partial\over\partial\tilde k}\tilde g_n(\tilde k)\nonu
&=&{\partial^n\over\partial\tilde\Phi^n}
\tilde k{\partial\over\partial\tilde k}\tilde U_{\tilde k}(\tilde\Phi),\\
&=&\left[n\left({d\over2}-1\right)-d\right]\tilde g_n
-\hbar{\Omega_d\over2(2\pi)^d}{\cal P}_n(\tilde G_2,\cdots,\tilde G_{n+2}),\nonumber
\eea
where
\be
k=\Lambda\tilde k,~~U=k^d\tilde U,~~\Phi=k^{{d\over2}-1}\tilde\Phi,~~
g_n=k^{n(1-{d\over2})+d}\tilde g_n,
\ee
and
\be
\tilde G_n={\tilde g_n\over1+\tilde g_2}.
\ee

\section{Scaling Regimes}
The $\phi^4$ model has two scaling regimes in the symmetrical phase, separated
by a crossover at the mass, $k_{cr}^2=g_2(k^2_{cr})$. For $k>>k_{cr}$
\be
G_n={g_n\over k^2}\left[1+O\left({g_2\over k^2}\right)\right],~~~~
\tilde G_n=\tilde g_n[1+O(\tilde g_2)],
\ee
and the evolution of the coupling constants is those of a massless theory with
the only scale of the cutoff. Below the crossover scale we have
\be
G_n={g_n\over g_2}\left[1+O\left({k^2\over g_2}\right)\right],~~~~
\tilde G_n={\tilde g_n\over\tilde g_2}[1+O(\tilde g_2^{-1})],
\ee
and the evolution comes to a halt due to the factor $k^d$ in right hand side of
(\ref{rgde}). This is as expected since a theory with mass gap has the quadratic
mass term as relevant operator and all non-Gaussian coupling constants are irrelevant.
The $\beta$-functions are always dominated by the highest order coupling constant that is
linear in the coupling constant in question. In fact, this term contains the lowest power
of $1/k^2$ at high energies. At the IR side this term is the most important because the
others contain the product of more non-Gaussian coupling constants which are supposed 
to be small.

Qualitatively new scaling proprieties can be established if the $\beta$-functions
are dominated by other terms or receive comparable contributions from different terms.
This is certainly the case when the restoring force,
\be
D(k)=k^2+g_2(k),~~~~\tilde D(\tilde k)=1+\tilde g_2(\tilde k),
\ee
is vanishing. Suppose that a theory specified at the cutoff by $U_\Lambda(\Phi)$
is in the symmetrical phase, i.e. $g_2(0)>0$ for the choice $\Phi_0=0$. The mass gap
decorrelates the field variables which are well separated in the space-time,
and the central limit theorem
asserts that $U_k(\Phi)$ approaches a quadratic form and $g_2(k)$ is 
monotonically increasing as $k\to0$. Perturbation expansion gives a more
detailed picture: $g_2(k)$ is a monotonically decreasing ($\beta_2<0$) and 
$D(k)>0$ is a monotonically
increasing function for all values of $k$. In order to find the new scaling laws
we need either a massless or symmetry broken-theory, where $D(k)$ reaches zero at
$k=0$ or at $k=k_0\not=0$, respectively. In the former case the Coleman-Weinberg 
mechanism \cite{cowe} generates a different vacuum where the theory manages
to develop a mass gap, thereby preserving the usual IR scaling laws. In the latter
case the vanishing of the inverse propagator, $D(k)$, indicates an instability
of the homogeneous vacuum in the effective theory with the cutoff $k\le k_0$
with respect to small fluctuations, the presence of the spinodal instability.
Thus new scaling laws might be found as the precursor of the spinodal phase 
separation. We should bear in mind that within the spinodal-unstable region
the saddle point is nontrivial, $\phi'_0\not=0$, and we need a different 
renormalization group equation.

\section{Renormalized Trajectory}
We want to follow the renormalized trajectory of the coupling constants 
\be
g_{2n}(k)={\partial^{2n}U_k(\Phi_0)\over\partial\Phi^{2n}}_{\big\vert\Phi_0=0},
\ee
in $d=4$, i.e. we seek for the solutions of the set of coupled equation 
\be
k{\partial\over\partial k}g_n(k)=\beta_n(g_1,\cdots,g_{n+2}),
\ee
where
\be
\beta_n=-\hbar{\Omega_4k^4\over2(2\pi)^4}
{\partial^n\over\partial\Phi^n}\ln\left[k^2+U_k''(\Phi)\right]_{\big\vert\Phi_0=0},
\ee
with the initial condition
\be
U_\Lambda(\Phi)=U_B(\Phi)
\ee
in the stable region, 
\be
k^2>k^2_0(\Phi)=-U''_{k_0(\Phi)}(\Phi).
\ee
We set $\hbar=1$ in the numerical work. We truncated the potential
\be
U_k(\Phi)=\sum_{n=1}^N{g_{2n}(k)\over (2n)!}\Phi^{2n},
\ee
and solved the resulting equations numerically. Special care is needed in the
vicinity of the spinodal line, $k\approx k_0(\Phi)$, where the $\beta$-functions
are the sum of large numbers with different sign. We used 3-nd and 4-th order
Runge-Kutta method with dynamically determined value of $\Dk$. Quadruple precision 
numbers were used when necessary to make sure that the roundoff errors in the 
$\beta$-functions were less than $10^{-8}$ times the actual value of the $\beta$-functions.

\begin{figure}
	\epsfxsize=10cm
	\centerline{\epsffile{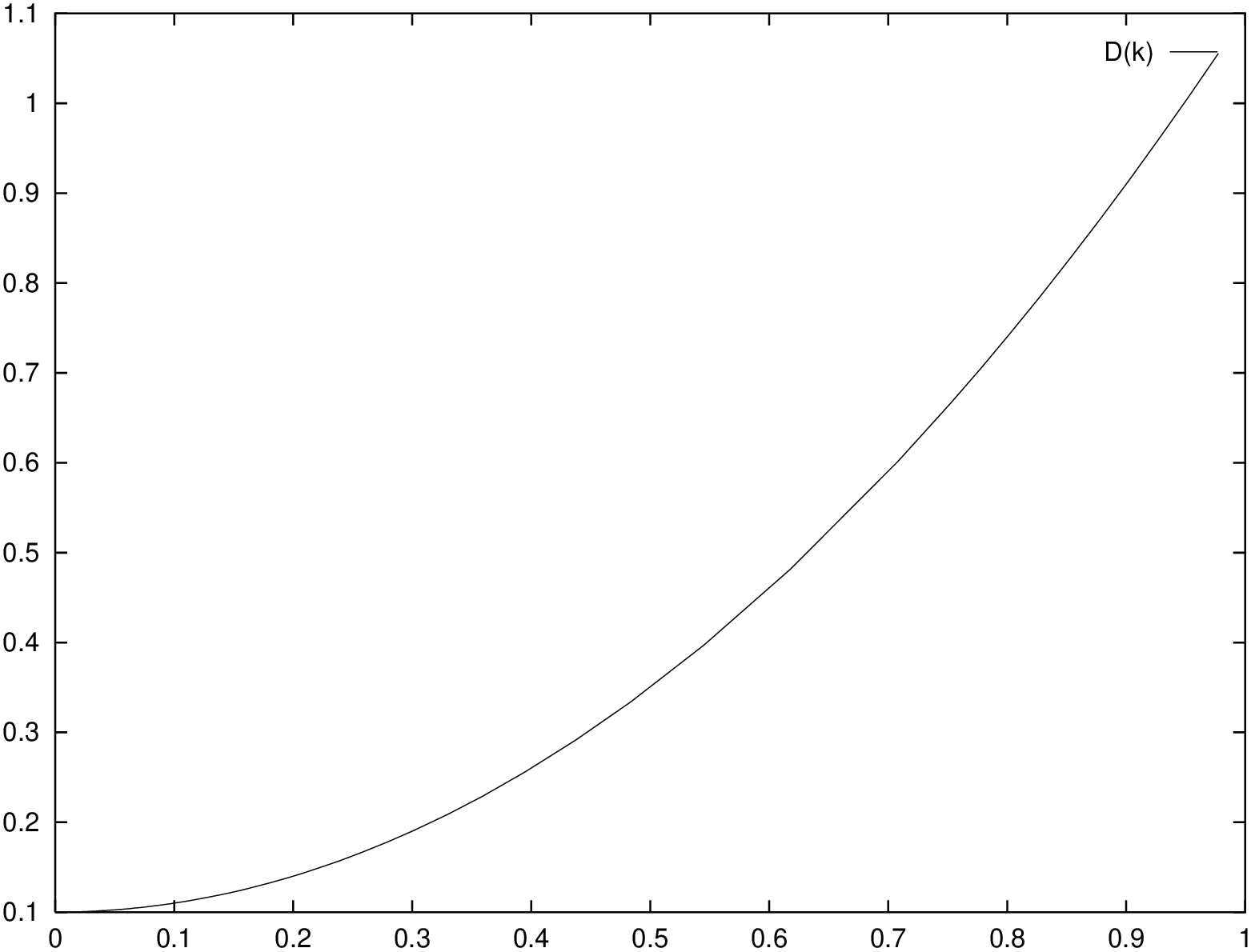}}
\centerline{(a)}
	\epsfxsize=10cm
	\centerline{\epsffile{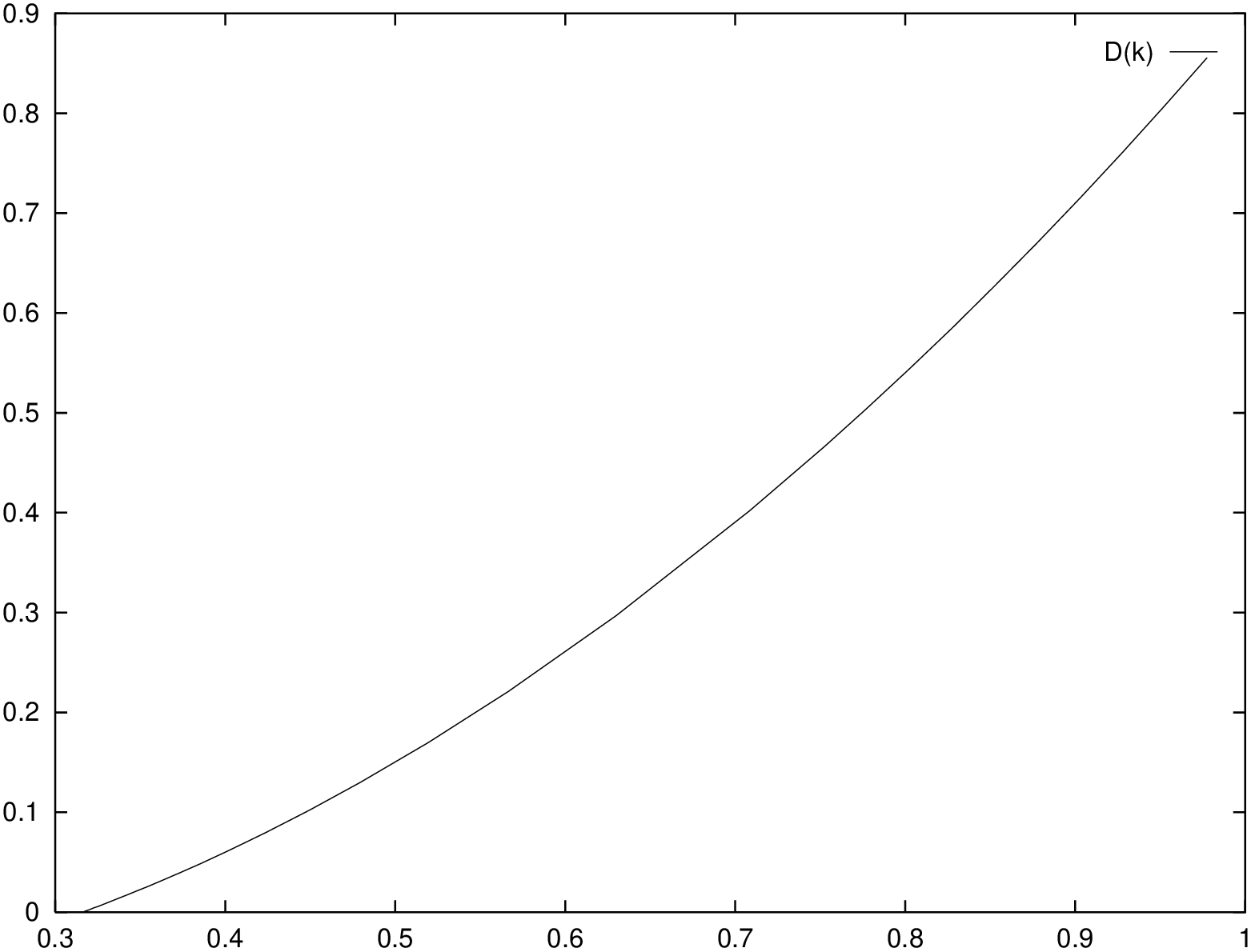}}
\centerline{(b)}
\caption{The evolution of the inverse propagator, $D(k)$ obtained by $2N=22$. 
(a) Symmetrical phase, $\tilde g_2(1)=0.1$, $\tilde g_4(1)=0.01$, 
(b) Symmetry broken phase, $\tilde g_2(1)=-0.1$, $\tilde g_4(1)=0.01$ In both 
cases $\tilde g_{2n}(1)=0$, for $n>2$.\label{inprop}}
\end{figure}

\begin{figure}
	\epsfxsize=10cm
	\centerline{\epsffile{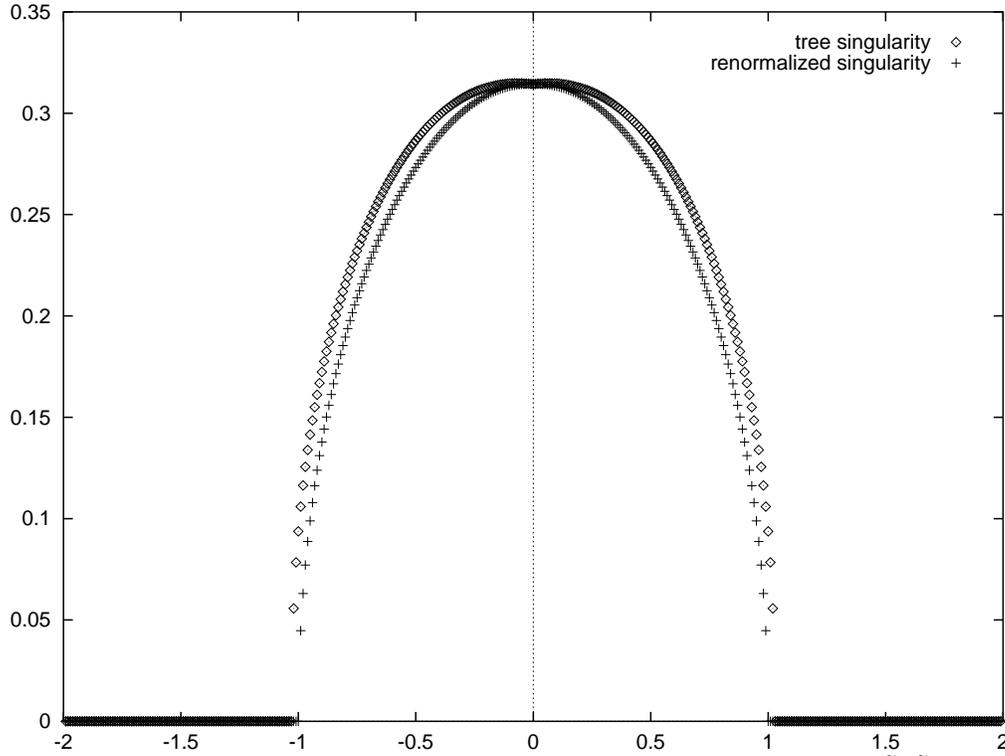}}
\caption{The line of singularity of the renormalized trajectory on the plane 
$(\tilde \Phi,\tilde k)$ for $\tilde g_2(1)=-.1$, $\tilde g_4(1)=0.2$. 
The diamond and the cross show the tree level and the renormalization
group results.\label{sing}}
\end{figure}

The behavior of the restoring force for the fluctuations, $D(k)$, was found
in qualitative agreement with the perturbation expansion, two typical cases are shown 
in Fig. \ref{inprop}. Such an agreement is expected in the UV scaling regime 
where the kinetic energy dominates the action. Our interest will be to see
the detailed behavior of the flow in the vicinity of the critical curve, 
$k>k_0(\Phi)$. The curves $k_0(\Phi)$ obtained from the tree-level solution and the
numerical integration of the renormalization group equations are shown in 
Fig. \ref{sing}. The quantum fluctuations help the disorder and drive the
saddle points to zero at the weakly unstable regime of the tree-level scaling
relations. As a result the curve obtained by solving the renormalization group
equations is inside of the spinodal-unstable region of the tree level solution,
except for the roundoff effect at small $k$.
It is remarkable that the radiative corrections are rather small.

The numerical integration of the renormalization group equations produces
oscillation for the coupling constants $g_n(k)$ for $n>4$, with increasing
amplitude as we approach the instability. We found two qualitatively 
different behaviors as far as the vicinity of the unstable line is
concerned. 

\begin{figure}
	\epsfxsize=10cm
	\centerline{\epsffile{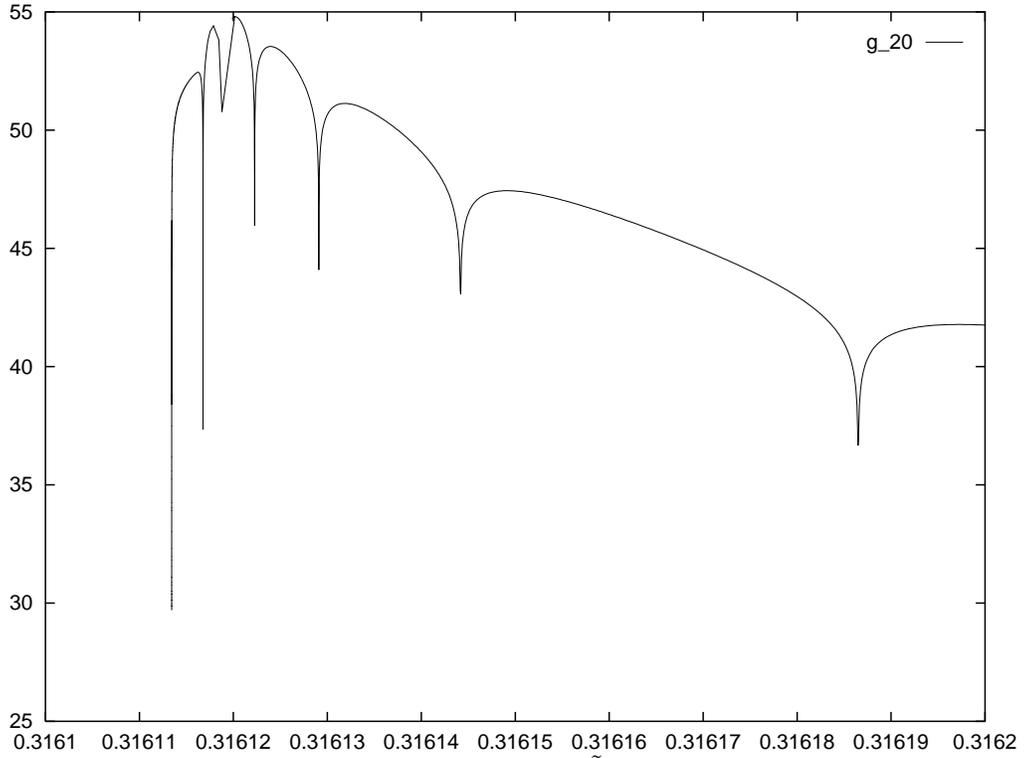}}
\caption{The evolution of the coupling constant $\tilde g_{20}(\tilde k)$ for
$2N=20$, with the initial conditions $\tilde g_2(1)=-0.1$, 
$\tilde g_4(1)=0.01$ and $\tilde g_{2n}(1)=0.0$ for $n=3,\cdots,10$.\label{evol10}}
\end{figure}

{\it Focusing.}
When the potential is truncated up to $2N=20$ then $g_4(k)$ stays positive 
and approaches zero. The values of the higher order coupling constants 
drop significantly after several, large-amplitude oscillations and approach
zero. This indicates that the blocking transformations has an attractive 
fixed point, 
\be
\tilde U_{\tilde k_0(0)}(\tilde\Phi)=-\hf\tilde\Phi^2.\label{fppot}
\ee
The coupling constants $g_n(k)$ for $n>4$ produce large fluctuations, but after
a while start to all fall and the quadratic potential, (\ref{fppot}) is
approached. It is found that $g_4(k)\to0^+$ and the higher order coupling
constants drop after undergoing large amplitude oscillations as $k\to k_0$. 
The evolution of $\ln|g_{20}(k)|$ is depicted in Fig. \ref{evol10}.
We find a cusp where the sign of $g_{20}(k)$ changes with finite 
altitude due to the finite resolution of the $k$ values.
Note that the negative coefficient of $\phi^2$ causes no problem with the
stability of the vacuum because the potential $U_k(\Phi)$ of the effective theory
recovers the perturbative form for large values of the field, far away from
the spinodal-unstable region. 

The approach to the Gaussian potential
can be made plausible by inspecting ${\cal P}_4$. By assuming that the
coupling constants remain finite at the critical line  we have
$g_4(k_0(0))=0$, since $\beta_4$ diverges otherwise. Once $g_4=0$ is
accepted, the vanishing of the higher order coupling constants is plausible.
The only finite parameter, $g_2(k_0(0))$, is fixed by the condition $D(k_0(0))=0$.
The existence of a single fixed point, that all finite set of coupling 
constants runs into (\ref{fppot}), can be called focusing. The potential 
turns out to be (\ref{fppot}) in the whole
unstable region that appears as a "fixed region" \cite{albapo}. 

\begin{figure}
	\epsfxsize=10cm
	\centerline{\epsffile{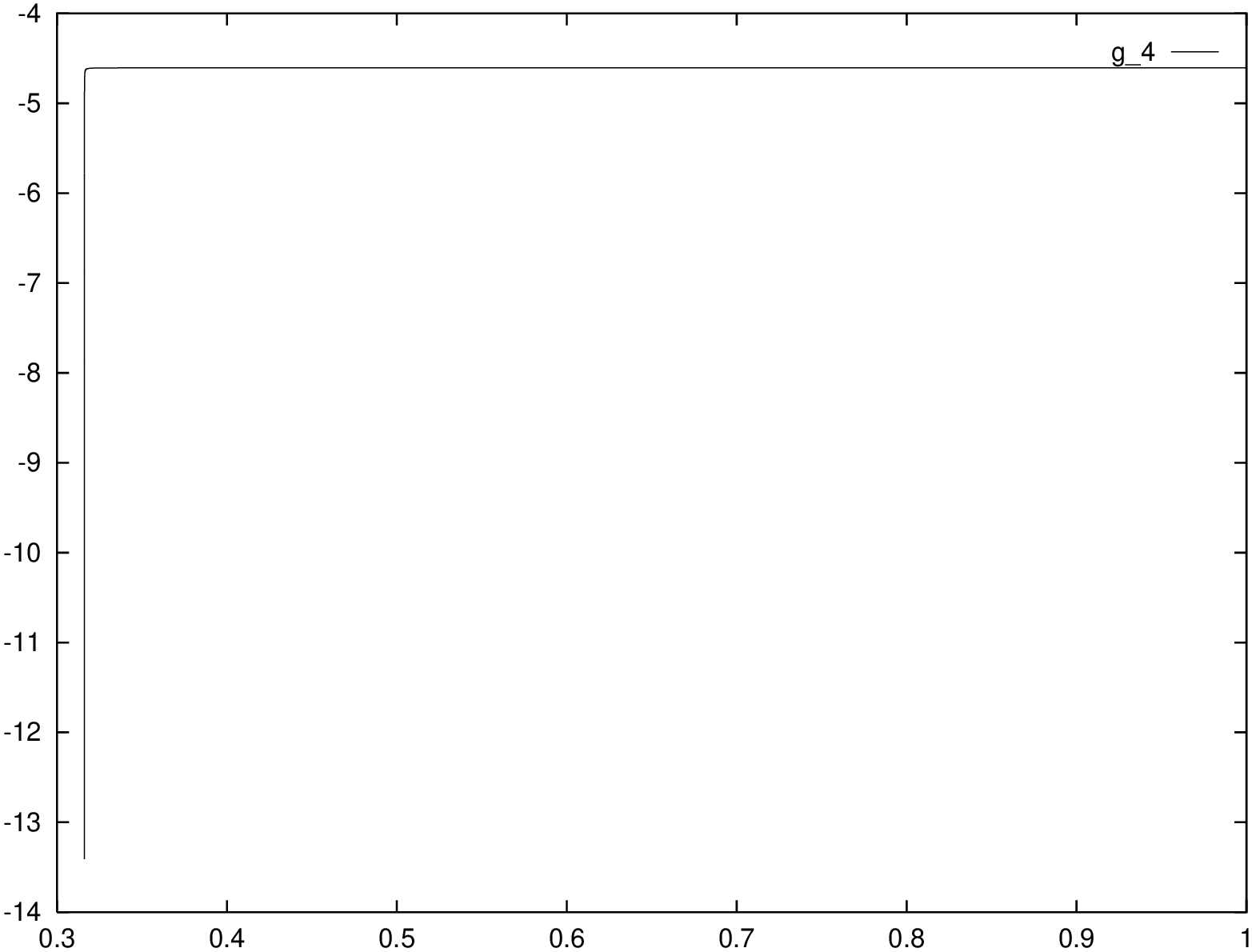}}
\centerline{(a)}
	\epsfxsize=10cm
	\centerline{\epsffile{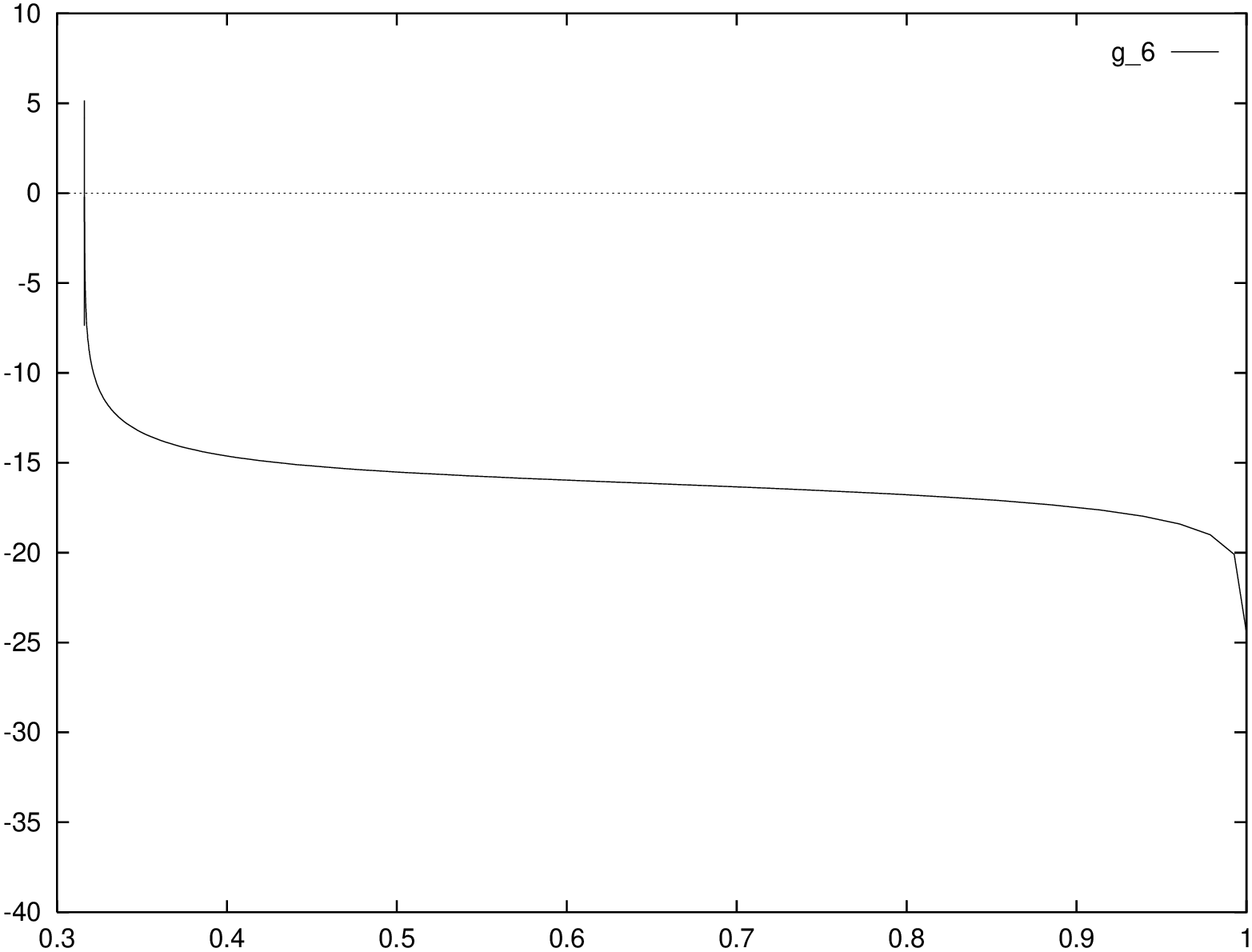}}
\centerline{(b)}
	\epsfxsize=10cm
	\epsfysize=9cm
	\centerline{\epsffile{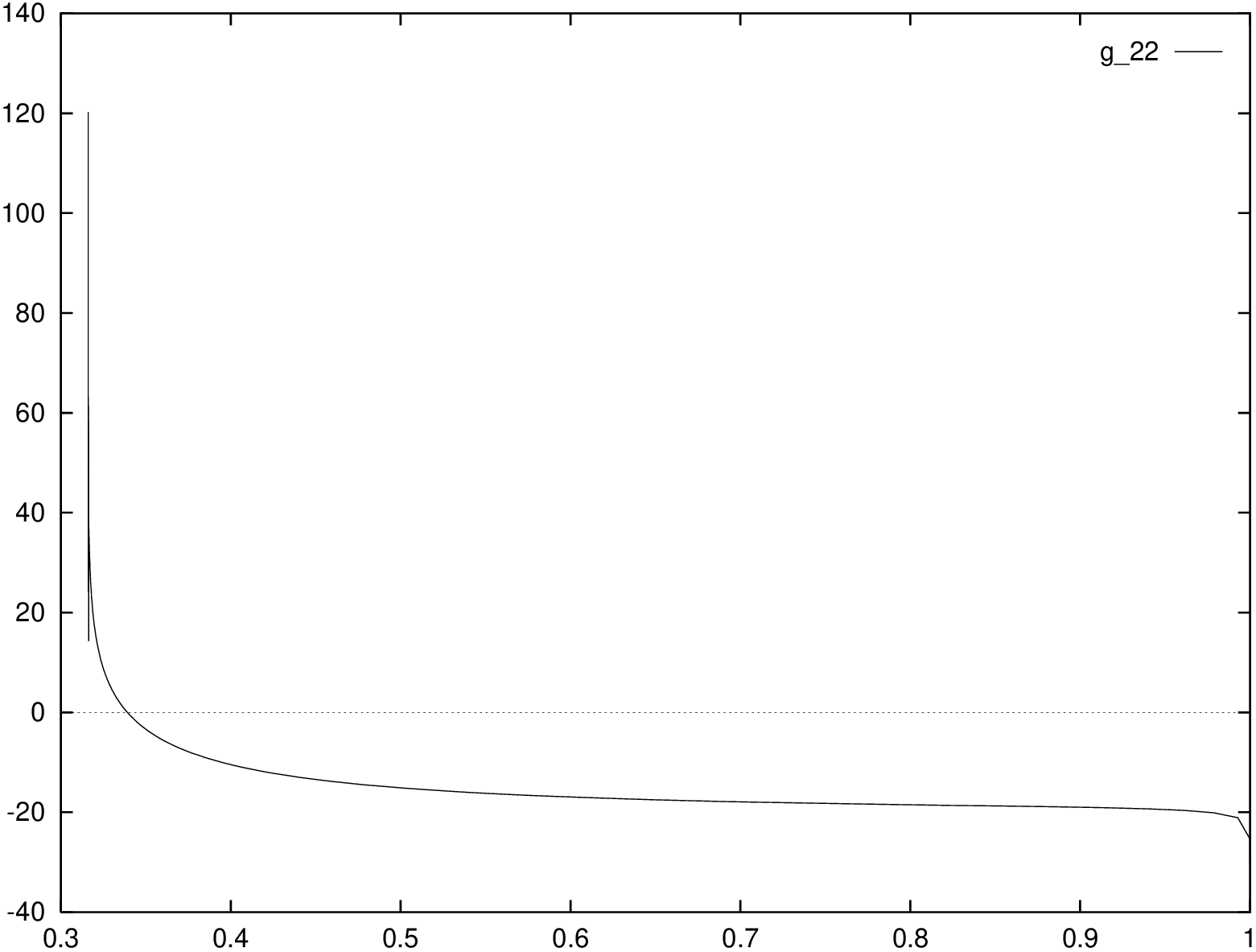}}
\centerline{(c)}
\caption{The evolution of: (a) $\ln|\tilde g_4(\tilde k)|$, 
(b) $\ln|\tilde g_6(\tilde k)|$ and 
(c) $\ln|\tilde g_{22}(\tilde k)|$ at $2N=22$, with
the initial conditions $\tilde g_2(1)=-0.1$, $\tilde g_4(1)=0.01$
and $\tilde g_{2n}(1)=0.0$ for $n=3,\cdots,11$. 
The coupling constants oscillate for $n>4$ with increasing amplitude and 
changing sign so $\ln|g_n(\tilde k)|$ is plotted.\label{evol11}}
\end{figure}

{\it Divergence.}
This fixed point turned out to be an artifact of the truncation of the
potential, a feature what has already been noted in other cases, as well, \cite{morrtr}.
When the truncation is made beyond $2N=20$ then the accumulation of the 
contributions of the higher order vertices in the $\beta$-functions 
make the term $G_6$ more important in $\beta_4$, which in turn helps 
$g_4$ to decrease faster with $k$ in the approach of the critical line. 
Once the sign of $g_4$, flips the further
decrease is not limited by zero and $g_4$ quickly approaches $-\infty$. It is not
then so surprising that all of the other coupling constants start to diverge at
the same time. The typical flow is depicted in Fig. \ref{evol11} for $2N=22$,
the further increase of $N$ makes no further qualitative changes in the flow.
The coupling constant undergoes oscillations with increasing amplitude
as $g_4(k)\to0^+$ and start to diverge as $g_4(k)$ flips sign. The
loop expansion naturally ceases to be applicable in the vicinity
of the critical line, and all we can say is that the modes with momentum
slightly above $k_0(\Phi)$ appear strongly coupled, and our solution is
no longer reliable. Though there is a
marked difference in the behavior of the renormalization group flow
for $2N<22$ and $2N\ge22$ but we should not forget that this difference
shows up after a strong coupling regime where the high order non-Gaussian
coupling constants develop extremely large values. So it is not clear if
the difference between two two cases in the vicinity of the unstable 
line is indeed so large.

\begin{figure}
	\epsfxsize=10cm
	\centerline{\epsffile{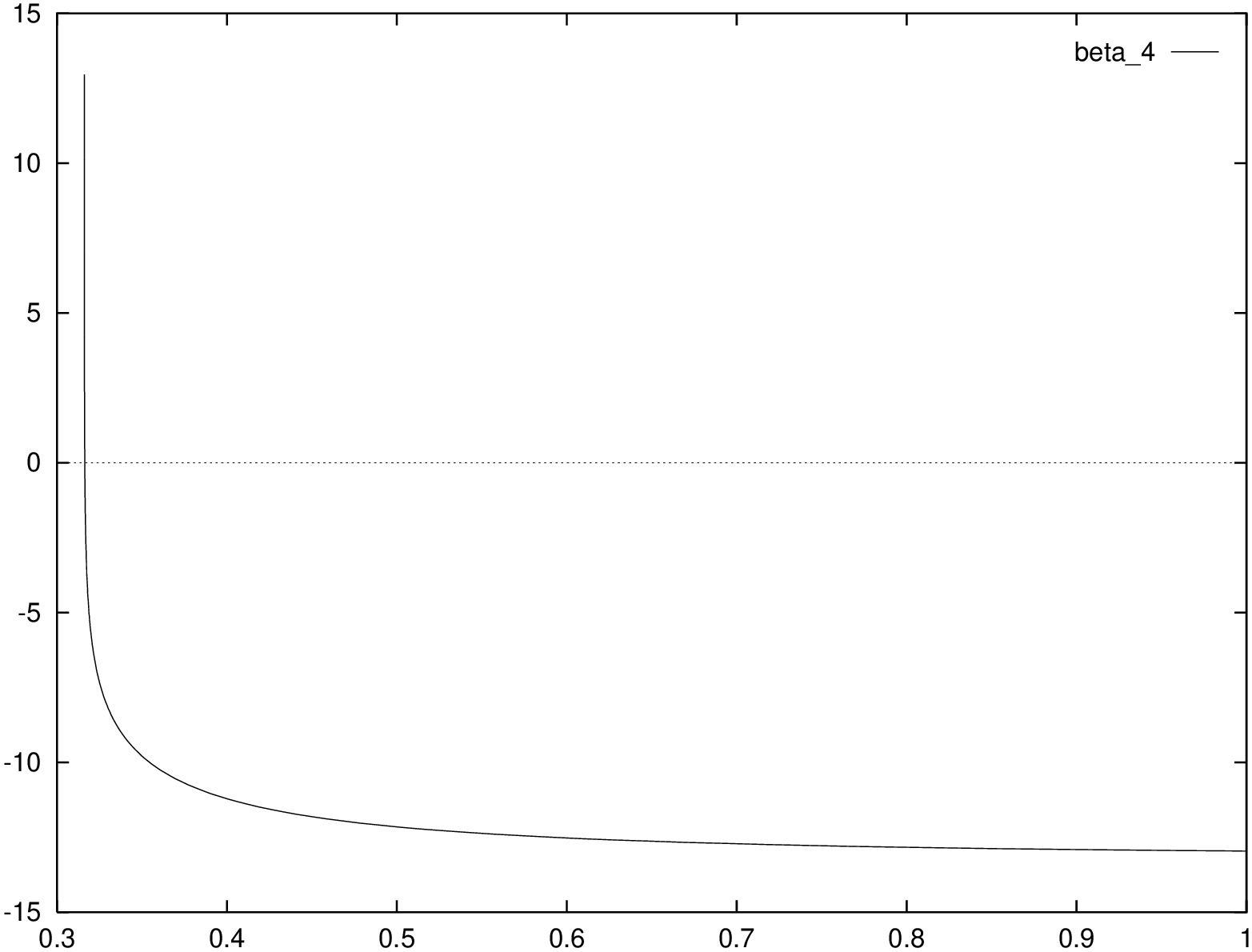}}
\centerline{(a)}
	\epsfxsize=10cm
	\centerline{\epsffile{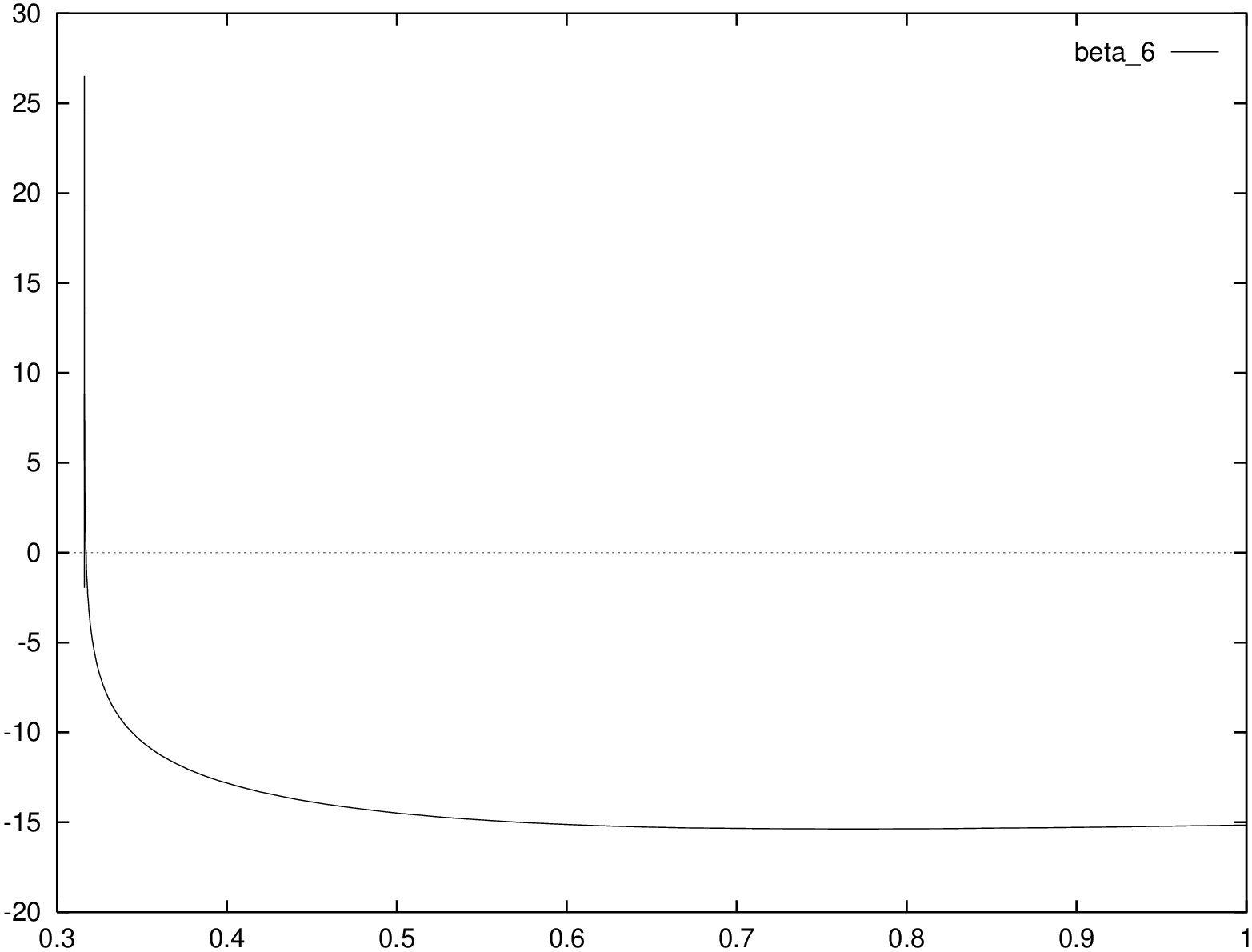}}
\centerline{(b)}
	\epsfxsize=10cm
	\centerline{\epsffile{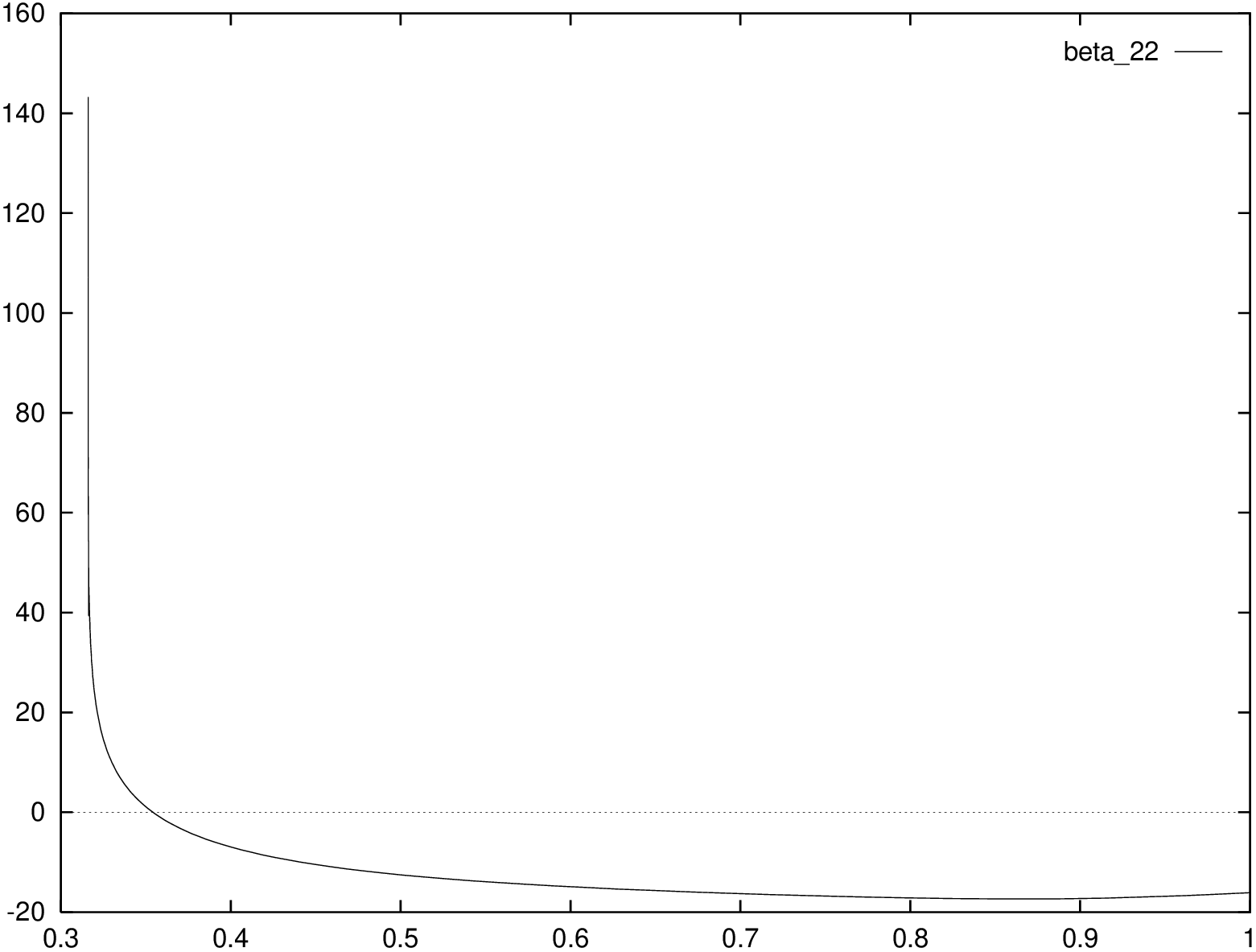}}
\centerline{(c)}
\caption{The evolution of 
$\ln|{\partial\tilde\beta_n(\tilde k)\over\partial\tilde g_6(1)}|$
with $2N=22$. (a): $n=4$, (b): $n=6$, (c): $n=22$.\label{fbfct}}
\end{figure}

{\it Universality.}
The increase of the coupling constants at $k\approx k_0(0)$ indicates the existence
of new relevant operator(s) in this scaling regime. One suspects that this
operator is nonlocal since the value of the cutoff is finite. Can this
operator modify the usual universality argument \cite{senben} ? According to
universality, the introduction of the irrelevant operators at the cutoff
modify only the scale parameter of the theory. The dimensionless quantities,
such as the $\beta$-functions, are supposed to be independent of the 
value of the irrelevant coupling constants at the cutoff. To verify this
scenario we computed
\be
{\partial\tilde\beta_n(\tilde k)\over\partial\tilde g_6(1)}\label{derbf}
\ee
numerically. The result, plotted in Fig. \ref{fbfct} shows clearly the 
coexistence of two different scaling regimes, the UV one where this quantity 
is suppressed and the precursor of the spinodal instability where we find
an increasing value. 

\begin{figure}
	\epsfxsize=10cm
	\centerline{\epsffile{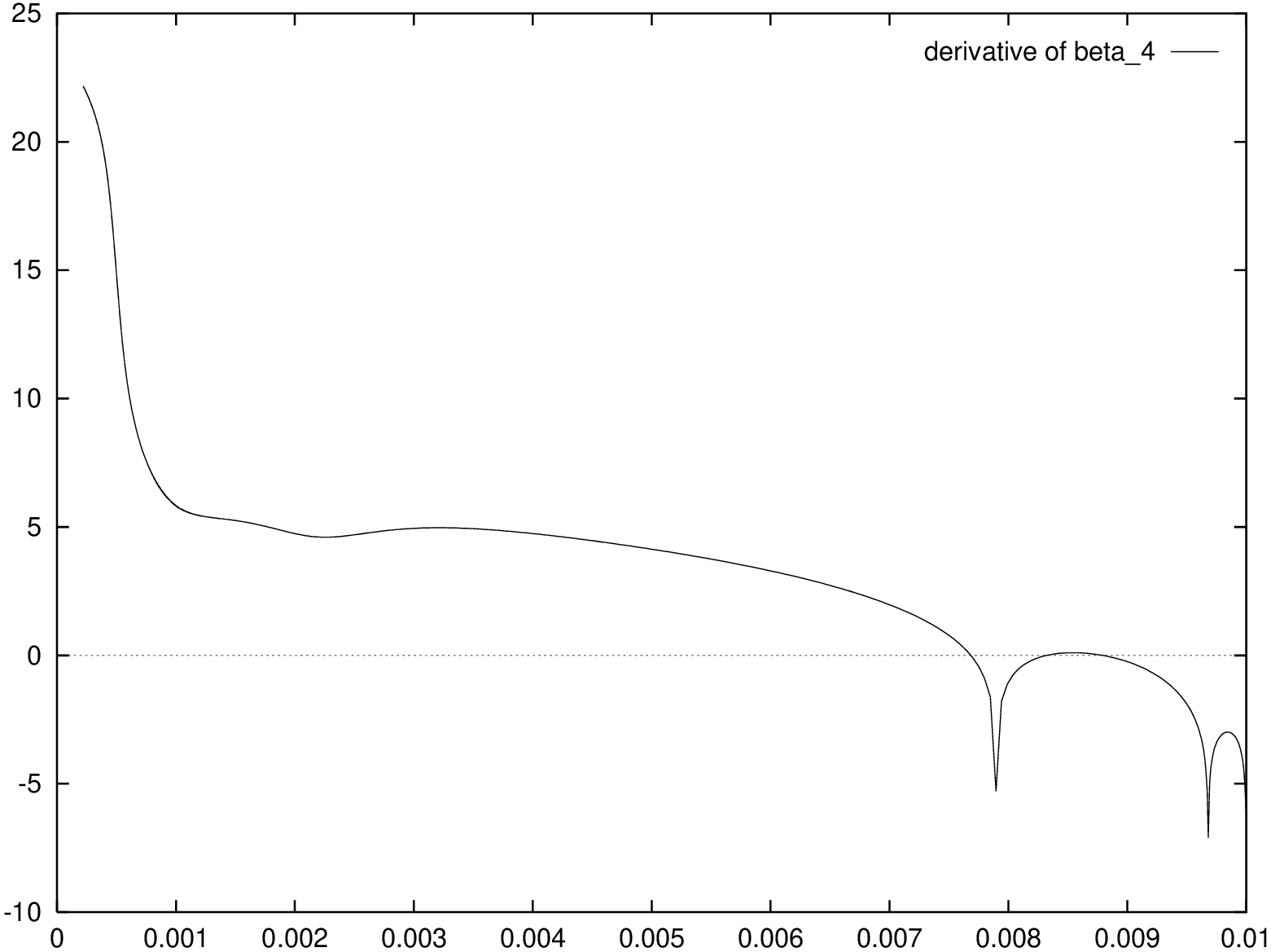}}
\centerline{(a)}
	\epsfxsize=10cm
	\centerline{\epsffile{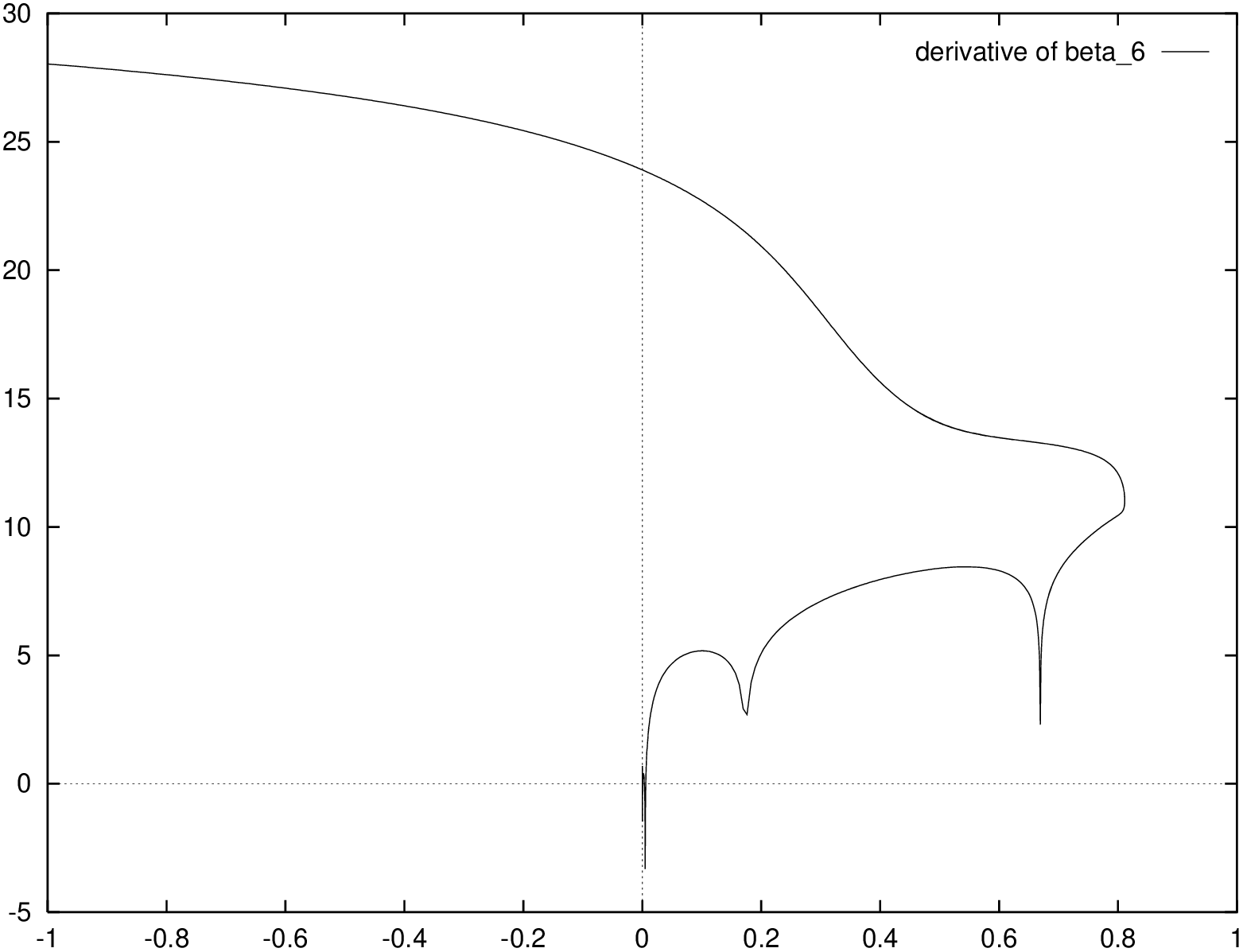}}
\centerline{(b)}
	\epsfxsize=10cm
	\centerline{\epsffile{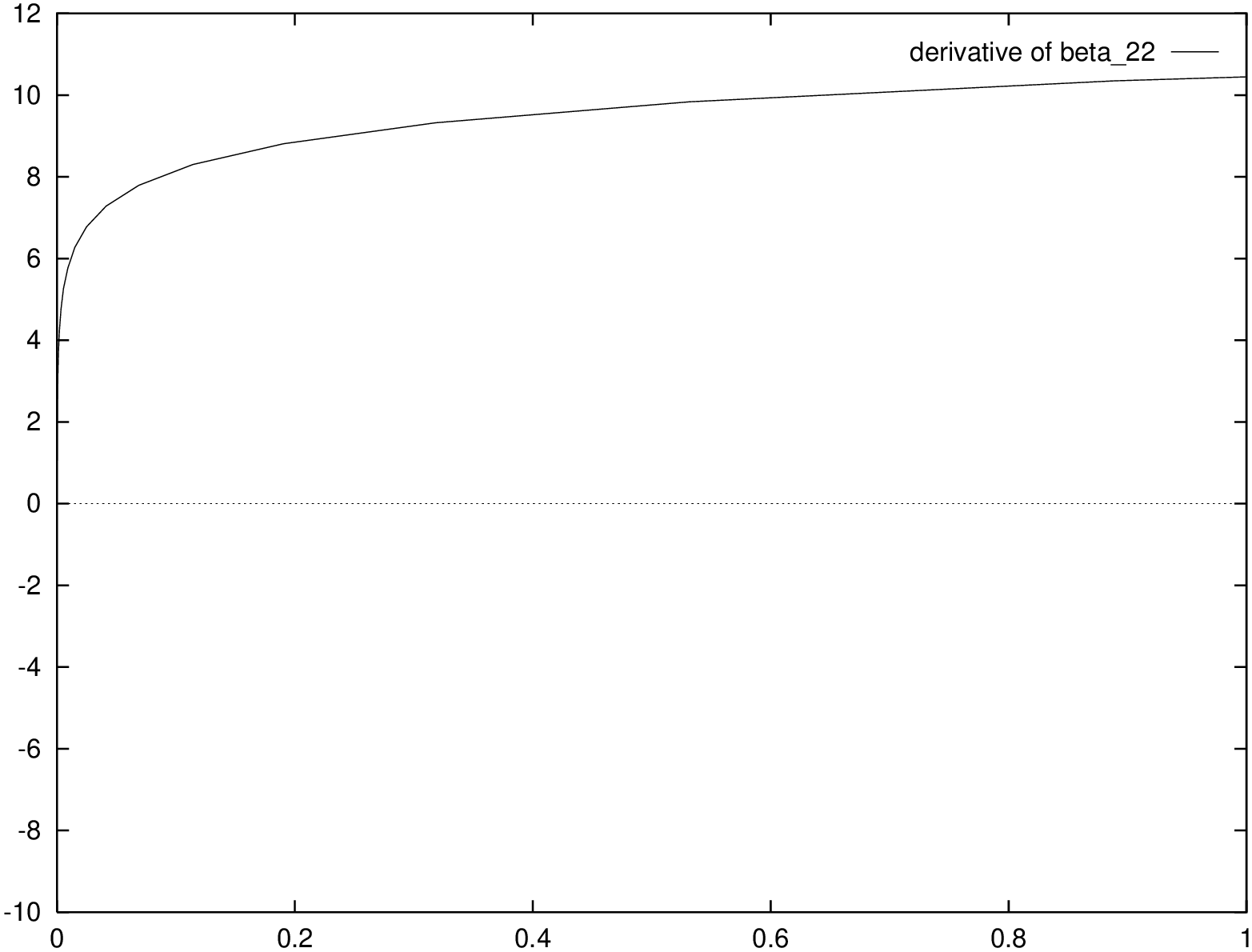}}
\centerline{(c)}
\caption{$\ln|\partial\tilde\beta_n(\tilde k)/\partial\tilde g_6(1)|$
plotted against $\tilde g_n(\tilde k)$ for (a): $n=4$, (b): $n=6$,  
(c): $n=22$, $0.317<\tilde k<1$. As the scale parameter $k$ decreases
we move towards left or right for (a) or (c), respectively. The derivative
of the beta function shows the onset of the low energy, nonuniversal
scaling already when the coupling constants are still weak enough to 
rely on the loop expansion.\label{bfgrat}}
\end{figure}

The comparison of Figs. \ref{evol11} and \ref{fbfct} contains an important lesson
confirmed by Fig. \ref{bfgrat} where (\ref{derbf}) is plotted as the function of 
the appropriate coupling constant. Namely, the violation of universality, the 
increase of (\ref{derbf}) takes already place when the coupling constants are
weak. It is reasonable to assume the ansatz
\be
{\partial\tilde\beta_n(\tilde k)\over\partial\tilde g_6(1)}
\approx F^2\left({\tilde k-\tilde k_0(0)\over\tilde k_0(0)}\right)
\tilde k_0^2(0),\label{ansa}
\ee
where the term $\tilde k_0^2(0)$ that is proportional to $\Lambda^{-2}$ represents the
suppression of the UV scaling laws, and the amplification effect of the 
instability is manifest in the behavior of $F(x)$, that is supposed to
diverge after large oscillations, as $x\to0$. For any finite value of 
the cutoff we can find a value of $k$ sufficiently close
to the instability where (\ref{ansa}) is unity,
\be
\Lambda\approx k_0(0)F\left({\tilde k-\tilde k_0(0)\over\tilde k_0(0)}\right),
\ee
and the cutoff effects of the nonrenormalizable coupling constant become visible
because the suppression of the UV scaling regime is compensated for by the 
amplification of the instability region. 

If the environment of the system, represented by the insertion of the constraint 
\be
\delta\left(\Phi-{1\over V}\int d^4x\phi(x)\right)
\ee
where $V$ is the four volume into the path integration is chosen in such a manner 
that the spinodal instability occurs, i.e. $\Phi$ is within the spinodal unstable 
region, then the amplification makes the effective coupling strength nonuniversal 
around the instability. In other words, the coupling constants of the effective 
theory for the system subject to this constraint can pick up the the values of certain
nonrenormalizable coupling constants at high energies, and allow us to
investigate the nonuniversal interactions at high energy if
the cutoff of the effective theory is brought close to $k=k_0(\Phi)$. 
By this method one could in principle increase the energy regime we can
access experimentally, and may get closer to the "last" important
scale, the onset of the asymptotic UV scaling of the TOE. 
This mechanism can be called a "renormalization group microscope" since the 
amplification offered by the instabilities is similar to the usual microscope,
except that it is achieved by the renormalization group flow in the space 
of the coupling constants. 

According to the numerical results the strength of the singularity and the value 
of (\ref{derbf}) approach zero at the critical line, as $\Phi$ is increased 
towards the edge of the unstable region $k=k_0(\Phi)$. This is the result
of the factor $k^d$ in the renormalization group equations, the decreasing 
entropy of the modes with weak restoring force as $k\to0$.

{\it Renormalized perturbation expansion.} In Statistical Physics one usually
follows the evolution of the bare coupling constants, $g_n(k)$, 
as the functions of the cutoff,
\be
k{\partial\over\partial k}g_n(k)=\beta_n^{(B)}(\{g\}),
\ee
where the explicit dependence on $k$ drops from the bare $\beta$-functions in the UV
region. We studied the evolution of these coupling constants in our work, too.

In Particle Physics one introduces the renormalized running coupling constants, 
$\lambda_n(\mu)$, $n=1,\cdots,n_r$, where $n_r$ is the number of renormalizable 
parameters in the theory. The irrelevant coupling constants are neglected
because the cutoff is sent sufficiently far from the observational energy, $\mu$.
The running coupling constants are defined by the help of the scattering amplitudes 
or Green functions and their scale dependence is described by the renormalization
group equations,
\be
\mu{\partial\over\partial\mu}\lambda_n(\mu)=\beta_n^{(R)}(\{\lambda\}),\label{evpp}
\ee
involving the renormalized $\beta$-functions.

So long as the perturbation expansion is applicable we can establish a 
one-to-one mapping between these schemes,
\be
g_n(k)=\lambda_n(k)+\cdots,
\ee
where the terms omitted are higher orders in the supposedly small coupling constants
or are $O(k/\Lambda)$.
In this manner the qualitative features of the renormalization group flow agree
in the two schemes, i.e. the running coupling constant becomes the bare 
one when the scale reaches the cutoff. How can we reconcile the importance of 
certain nonrenormalizable coupling constants at low energy with the 
evolution of the running coupling constants, $\lambda_n(\mu)$, which are introduced
to keep track of the renormalizable coupling constants only ?

The independence of the theory from the observational scale,
\be
\mu{d\over d\mu}\Gamma_B(\{\lambda_B\},\Lambda)=0,
\ee
is reached by the readjustment of the coupling constants, $\lambda_n(\mu)$.
According to the multiplicative renormalization scheme the number of
adjustable parameters is just $n_r+1$. In particular, one chooses $n_r+1$ 
independent observables, $\Gamma^{(m)}$, $m=1,\cdots,n_r+1$,
\be
\Gamma_B^{(m)}(\{\lambda_B\},\Lambda)=Z^{\ell_m}
\left((\{\lambda(\mu)\},{\Lambda\over\mu}\right)
\Gamma_R^{(m)}(\{\lambda(\mu)\},\mu),
\ee
and imposes
\bea
0&=&\mu{d\over d\mu}\left[Z^{\ell_m}\left((\{\lambda(\mu)\},{\Lambda\over\mu}\right)
\Gamma_R^{(m)}(\{\lambda(\mu)\},\mu)\right]\\
&=&\left(\mu{\partial\over\partial\mu}+\beta_n^{(R)}(\{g\}){\partial\over\partial g_n}
+\ell_m\gamma^{(R)}(\{g\})\right)\Gamma_R^{(m)}(\{\lambda(\mu)\},\mu).
\eea
The renormalized renormalization group functions, $\gamma^{(R)}$
and $\beta_n^{(R)}(\{\lambda\})$ can be found by inverting 
\bea
0&=&\mu{d\over d\mu}\left[Z^{\ell_m}\left((\{\lambda(\mu)\},{\Lambda\over\mu}\right)
\Gamma_R^{(m)}(\{\lambda(\mu)\},\mu)\right]\\
&=&\left(\mu{\partial\over\partial\mu}+\beta_n^{(R)}(\{g\}){\partial\over\partial g_n}
+\ell_m\gamma^{(R)}(\{g\})\right)\Gamma_R^{(m)}(\{\lambda(\mu)\},\mu)\label{rgeqin}
\eea
for the $\gamma^{(R)}$ and the $\beta^{(R)}$ functions.  The perturbative renormalization
assures that these functions
are well defined, i.e. are independent of the choice of the observables $\{\Gamma\}$. 

Observe that there are two steps in this procedure which prevent the detection
of a hidden coupling constant. One is that we have already 
committed ourselves to use
as many coupling constants to compensate the modification of $\mu$, as many renormalizable
operators are in the system. The second is that the cutoff is formally removed from the 
renormalized perturbation expansion. The hidden coupling constant represents a 
coupling between the ultraviolet and the infrared  modes that can be seen by
keeping the UV and IR cutoffs finite. When the UV cutoff is removed
then the contributions $O(\mu/\Lambda)$ are ignored. The renormalization
group used in this work resums these contributions which yield the new nontrivial scaling laws.
Thus the observables obtained by the improved renormalized 
perturbation expansion using the renormalization group for $n_r$ coupling constants will, by 
construction, never show any indication of the eventual, hidden parameters.
What we have found is that our partial resummation of the perturbation 
expansion, which keeps track of nonrenormalizable operators as well, 
indicates the presence of hidden coupling constants and suggests that the 
low energy dynamics is parametrized by more than $n_r$ coupling constants. 
Consequently
more than $n_r+1$ observables must be used in (\ref{rgeqin}) to obtain the 
evolution of the parameters that can compensate the change of the observational 
scale. 

In order to find out the number of real parameters, we need a
controllable method to study the low-energy scaling behavior. Due to the
limitation of the gradient expansion based on local operators we can not
at the present stage clarify this point.

\section{Summary}
The renormalization group is traditionally used to follow the scale
behavior in the vicinity of a fixed point of the blocking transformation.
We showed in the case of the $\phi^4$ model that, after paying the price 
of following the mixing of a large number of operators during the blocking,
the investigation of the manner by which the different scaling regimes give
rise to each other is feasible. 

A new finite energy scaling regime of the $\phi^4$ model with spontaneously
broken symmetry is generated by the spinodal instability. It was found,
by the numerical integration of the Wegner-Houghton equation in the
local-potential approximation, that the spinodal instability generates
new relevant operators, and may undo the suppression of the nonrenormalizable 
operators at the UV scaling regime. A nonrenormalizable operator gives rise to
a hidden coupling constant if the operator in question is relevant in the
low-energy scaling regime, and the initial, high energy value of its
coupling constant influences the low energy physics.
This raises the possibility of the eventual use of this instability as 
a renormalization group microscope to detect the nonuniversal physics at 
high-energy by going sufficiently close to the unstable region.

The instability studied in this work appears in the Euclidean effective-theory
at finite-momentum scales. It is conjectured that the simplest manner
to observe the effects of the instability is in the framework of the
dynamical renormalization group, applied for large amplitude fluctuations.
The saddle-point structure of the effective theory, for the real time 
dependence, is needed to make more definite proposal, concerning the whereabouts 
of nonuniversal phenomena.  
The investigation of the renormalization group equation with nontrivial
condensate within the spinodal phase separation, and the search for the
effects of the instability in the vacuum with Goldstone modes, is in
progress \cite{albapo}.

\begin{table}
\begin{center}
\begin{tabular}{@{}*{4}{|l|l||l|}}
\hline    
U.V.&I.R.&Fig. \ref{qedccp}&example\cr 
\hline    
\hline
relevant&relevant&(a)&$\tilde m_e\Lambda\bar\psi_e\psi_e$\cr
relevant&irrelevant&(b)&$\tilde m_\mu\Lambda\bar\psi_\mu\psi_\mu$\cr
irrelevant&relevant&(c)&$\tilde G\Lambda^{-2}(\bar\psi_e\psi_e)^2$\cr
irrelevant&irrelevant&(d)&$\tilde c\Lambda^{-5}(\bar\psi_e\psi_e)^3$\cr
\hline
\end{tabular}
\caption{The four classes of the coupling constants in QED.\label{qedcc}}
\end{center}
\end{table}

\acknowledgments
We thank Daniel Boyanowski, Pino Falci, Rosario Fazio, Peter Hasenfratz, 
David Jasnow and Dario Zappal\`a for useful discussions.

\appendix
\section{Wegner-Houghton Equations}
The heuristic derivation of the renormalization group equation (\ref{rgde})
indicates that the higher loop contributions to the equation are suppressed.
This is not obvious from the derivation presented above because by placing the
system into a finite quantization box the spectrum of the momentum becomes
discrete and we may eliminate the modes ony-by-one. What is the small
parameter in this case?

The expansion of the action around the constant background $\Phi$ in powers of 
the Fourier components of fluctuations $\phi'$ is
\be
S_k[\Phi+\phi']=S_k[\Phi]+
\sum_p\phi'_p\frac{\partial S_k}{\partial\phi_p}|_{\phi}+
\hf\sum_{p,q}\phi'_p\phi'_q\frac{\partial^2 S_k}
{\partial\phi_p\partial\phi_q}|_{\phi}+...
\ee
We have
\bea
\frac{\partial S_k}
{\partial\phi_p}|_{\phi}&=&U_k^{(1)}(\Phi)L^d\delta(p)\nonumber\\
\frac{\partial^2 S_k}{\partial\phi_{p_1}\partial\phi_{p_2}}|_{\phi}&=&
\left[U_k^{(2)}(\Phi)+p^2\right]L^d\delta(p_1+p_2)\nonumber\\
\frac{\partial^n S_k}{\partial\phi_{p_1}...\partial\phi_{p_n}}|_{\phi}&=&
U_k^{(n)}(\Phi)L^d\delta(p_1+...+p_2)
\eea
where the subscript stands for the derivatives with respect to $\Phi$.
Since $k-\Dk<|p|<k$, the first derivative of the action does not contribute,
\be\label{devact}
S_k[\Phi+\phi']=S_k[\Phi]+
\frac{L^d}{2}\sum_p\phi'_p\phi'_{-p}\left[U_k^{(2)}(\Phi)+p^2\right]+...
\ee
The minimum value of $\Dk$ being $2\pi/L$ where $L$ is the length of the quantization
box, the number of modes to eliminate in the shell $k-\Dk<|p|<k$ is
\be\label{nd}
{\cal N}_d=\frac{\Omega_d k^{d-1}\Dk}{(2\pi/L)^d}=
\frac{\Omega_d}{2\pi}\left(\frac{k}{2\pi}\right)^{d-1}L^d\Dk
\ee
where $\Omega_d$ is the solid angle in dimension $d$.
The integration over degrees of freedom $\phi'$ will be done after the 
expansion of the exponential around the free action. The only terms contributing
in the integration are those for which the Fourier components of $\phi'$ are 
combined in pairs $\phi'_p\phi'_{-p}$,
\bea
e^{-\frac{L^d}{\hbar}(U_{k-\Dk}(\Phi)-U_k(\Phi))}&=&
\int{\cal D}[\phi']e^{-\frac{1}{2\hbar}
\sum_p\left[k^2+U_k^{(2)}(\Phi)\right]\phi'_p\phi'_{-p}}\\
&&\times\left(1-\frac{U_k^{(4)}(\Phi)}{2L^d\hbar}\sum_{p,q}\phi'_p\phi'_{-p}
\phi'_p\phi'_{-q}+...\right)\nonumber
\eea
where ${\cal D}[\phi']=\prod_{k-\Dk<|p|<k}d{\cal R}e(\phi'_p)d{\cal I}m(\phi'_p)$.
The Gaussian integrations lead to
\bea\label{devres}
e^{-\frac{L^d}{\hbar}(U_{k-\Dk}(\Phi)-U_k(\Phi))}&=&
\left(\frac{\hbar\pi}{k^2+U_k^{(2)}(\Phi)}\right)^{{\cal N}_d/2}\\
&&\times\left[1-\frac{\hbar}{L^d}\frac{U_k^{(4)}(\Phi)}{[k^2+U_k^{(2)}(\phi)]^2}
{\cal N}_d\left(\frac{{\cal N}_d}{2}+1\right)+...\right].\nonumber
\eea

We now introduce the small variable,
\be\label{hypdev}
\hbar^{n-1}\frac{|U_k^{(2n)}(\phi)|}{[k^2+U_k^{(2)}(\phi)]^2}\frac{({\cal N}_d)^n}{(L^d)^{n-1}}<< 1 
\ee
what is proportional with the ratio of the number of degrees of freedom eliminated
in a blocking to those left in the effective theory.
We must keep in mind that first we have to take the thermodynamical limit, $L\to\infty$
and after $\Dk\to0$ to make sure that the higher loop contributions are small.
In principle $\Dk$ has a lower bound $2\pi/L$ but we can imagine that we make
an interpolation on the renormalized trajectory so that the value of $\Dk$ we use 
to derive the renormalization group equation is as small as we wish. In this manner 
$\Dk$ and $L$ are independent and the small parameter is $\Dk/M$. If we want to 
keep the lower bound for $\Dk$ nonvanishing then we must assume that the derivatives 
of the potential are small enough according to (\ref{hypdev}) and therefore the
existence of another small parameters hidden in the potential.

By assuming that (\ref{hypdev}) is valid and taking the logarithm of both sides 
of (\ref{devres}) we obtain
\bea
U_k(\Phi)-U_{k-\Dk}(\Phi)&=&C(k)-\Dk\frac{\hbar\Omega_d k^{d-1}}{2(2\pi)^d}
\ln\left(k^2+U_k^{(2)}(\Phi)\right)\\
&+&\frac{(\Dk)^2}{2}\left(\frac{\hbar\Omega_dk^{d-1}}{(2\pi)^d}\right)^2
\left[\frac{U_k^{(4)}(\Phi)}{(k^2+U_k^{(2)}(\Phi))^2}\right]+...\nonumber
\eea
what yields finally to the Wegner-Houghton equation 
\be
\partial_kU_k(\Phi)=-\frac{\hbar\Omega_d k^{d-1}}{2(2\pi)^d}
\ln\left(\frac{k^2+U_k^{(2)}(\Phi)}{k^2+U_k^{(2)}(0)}\right)
\ee
where the denominator in the logarithm function was inserted to cancel
the potential at $\Phi=0$.

\end{document}